\documentclass{IEEEoj}

\usepackage[mathscr]{euscript}
\let\euscr\mathscr \let\mathscr\relax
\usepackage[scr]{rsfso}
\usepackage[active]{srcltx} 
\usepackage[cmex10]{amsmath}
\usepackage{amsfonts}
\usepackage[final]{graphics}
\usepackage[final]{graphicx}
\usepackage{amsbsy}
\usepackage{amsbsy}
\usepackage{amssymb}
\usepackage{url}
\usepackage{enumerate}
\usepackage{cite}
\usepackage{psfrag}
\usepackage{color}
\usepackage{euscript}
\usepackage[utf8]{inputenc}
\usepackage{algorithmic}
\usepackage[ruled,vlined]{algorithm2e}
\usepackage{xcolor}
\usepackage{soul}
\usepackage{tikz}
\usepackage{textcomp}
\usepackage{lipsum}
\usepackage[T1]{fontenc}
\usepackage{textcomp}
\usepackage{multirow}
\usepackage{anyfontsize}

\def\BibTeX{{\rm B\kern-.05em{\sc i\kern-.025em b}\kern-.08em
    T\kern-.1667em\lower.7ex\hbox{E}\kern-.125emX}}
\AtBeginDocument{\definecolor{ojcolor}{cmyk}{0.93,0.59,0.15,0.02}}
\def\OJlogo{\vspace{-4pt}\hskip-4pt\includegraphics[height=18pt]{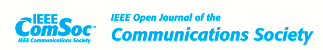}}
\newcommand{\bm}[1]{{\mathbf{#1}}}
\newcommand{\Es}{{\mathbb{E}}}          

\newcommand{\diag}{{\text{diag}}}

\newcommand{\br}{\bm r}

\newcommand{\Cset}{\mathbb{C}}
\newcommand{\Rset}{\mathbb{R}}

\newcommand{\eqdef}{\triangleq}

\newcommand{\herm}{\text{H}}
\newcommand{\trasp}{\text{T}}

\newcommand{\rate}{\EuScript{R}}

\def\bdm#1\edm{\begin{displaymath}#1\end{displaymath}}
\def\be#1\ee{\begin{equation}#1\end{equation}}
\def\barr#1\earr{\begin{align}#1\end{align}}

\newcommand{\IeeeTIT}{{\em IEEE Trans.\ Inf. Theory\/}}

\newcommand{\IeeeTCOMM}{{\em IEEE Trans.\ Commun.\/}}

\newcommand{\IeeeWCOMMLETT}{{\em IEEE Wireless Commun.\ Lett.\/}}

\newcommand{\IeeeJSAC}{{\em IEEE J.\ Select.\ Areas Commun.\/}}

\newcommand{\IeeeTAP}{{\em IEEE Trans.\ Antennas Propag.\/}}

\usepackage{tikz}

\newcommand\copyrighttext{%
  \footnotesize \textcopyright \the\year{} IEEE. Personal use of this material is permitted. Permission from IEEE must be obtained for all other uses, including reprinting/republishing this material for advertising or promotional purposes, collecting new collected works for resale or redistribution to servers or lists, or reuse of any copyrighted component of this work in other works.}

\newcommand\acceptedtext{%
  \footnotesize This article has been accepted for publication in a future issue of this journal,
  but has not been fully edited. Content may change prior to final publication. \\
  Citation information: DOI 10.1109/OJCOMS.2025.3526126, IEEE Open Journal of Communications Society.}
  
\newcommand\acceptednotice{%
\begin{tikzpicture}[remember picture,overlay]
\node[anchor=north,yshift=12pt,xshift=72pt] at (current page.north) {%
\begin{minipage}{\textwidth}
\center \acceptedtext
\end{minipage}};
\end{tikzpicture}%
}

\begin{document}

\receiveddate{12 December, 2024}
\reviseddate{}
\accepteddate{2 January, 2025}
\publisheddate{XX Month, 2025}
\doiinfo{10.1109/OJCOMS.2025.3526126}

\title{Design of stacked intelligent metasurfaces 
with reconfigurable amplitude and phase
for multiuser downlink beamforming 
}
\author{DONATELLA~DARSENA\authorrefmark{1} (SENIOR MEMBER, IEEE),
FRANCESCO~VERDE\authorrefmark{1} (SENIOR MEMBER, IEEE),
IVAN~IUDICE\authorrefmark{2}, AND
VINCENZO~GALDI\authorrefmark{3} (FELLOW, IEEE)
}
\affil{Department of Electrical Engineering and
Information Technology, University Federico II, I-80125 Naples, Italy}
\affil{Security Unit, Italian Aerospace Research Centre (CIRA), I-81043 Capua, Italy}
\affil{Department of Engineering, University of Sannio,  
I-82100 Benevento, Italy}
\corresp{CORRESPONDING AUTHOR: F.~VERDE (e-mail: f.verde@unina.it).}
\authornote{This work was partially supported by 
the European Union-Next Generation EU under the Italian
National Recovery and Resilience Plan (NRRP), Mission 4,
Component 2, Investment 1.3, CUP E63C22002040007, partnership
on ``Telecommunications of the Future" (PE00000001
- program ``RESTART").}
\markboth{Designs of stacked intelligent metasurfaces 
with reconfigurable amplitude and phase
for multiuser downlink beamforming}{Darsena \textit{et al.}}

\begin{abstract}

A novel technology based on stacked intelligent
metasurfaces (SIM) has recently emerged. This platform involves 
cascading multiple metasurfaces, each acting as a digitally programmable physical layer within 
a diffractive neural network. SIM enable the implementation of signal-processing transformations directly in the electromagnetic wave domain,
eliminating the need for expensive, high-precision, and power-intensive digital platforms. 
However, existing studies employing SIM in wireless communication 
applications  rely solely on nearly passive structures that  
control only the phase of the meta-atoms in each layer. 
In this study, we propose a SIM-aided downlink multiuser transmission scheme, 
where the SIM at the base station (BS) end is designed by combining 
nearly passive layers with phase-only reconfiguration capabilities 
and active layers integrated with amplifier chips to enable amplitude control.
Our optimal design aims at maximizing  the sum rate for 
the best group of users by jointly 
optimizing the transmit power allocation at the BS and
the wave-based beamforming at the SIM.
In addition to the standard sum-power constraint at the BS,
our optimization framework includes
two additional constraints: (i) 
a per-stream power preserving constraint to prevent propagation losses across the SIM, and
(ii) an amplitude constraint 
to account for power limitations for
each active layer.
To further reduce the complexity of the optimal beamforming 
solution, we explore a simple yet suboptimal 
zero-forcing (ZF) beamforming design,  where the wave-based
transformation implemented by the SIM is selected to eliminate 
interference among user streams.
Finally, extensive Monte Carlo simulations demonstrate that incorporating
both nearly passive and active layers within the SIM significantly enhances 
capacity compared to previously reported phase-only
coding SIM. Additionally, the numerical results reveal that 
low-complexity ZF beamforming approaches optimality in terms of 
maximum sum rate even for a relatively small number of users.

\end{abstract}


\maketitle

\acceptednotice

\section{Introduction}

\IEEEPARstart{S}{ignal} processing and communication digital technologies
have evolved over decades, resulting in 
increasingly sophisticated platforms and algorithms that are
power-hungry and require complex initialization procedures.  
For instance, within machine learning (ML) techniques, 
deep learning (DL) has emerged as one of the fastest-growing approaches \cite{Lecum.2015}. 
However, this advancement has led to a significant increase in computing power requirements, 
resulting in high energy consumption and extended training times \cite{Caulfield.2010}.
On the other hand, massive {\em multiple-input multiple-output (MIMO)} offers a promising
solution to manage the exponential growth in data traffic and 
meet the increasing demands for communication service quality \cite{Marzetta-book}.
However, massive MIMO systems come with challenges: they require high-resolution 
digital-to-analog converters (DAC) and analog-to-digital converters (ADC) for digital beamforming,
which raises hardware costs. Additionally, the need for numerous radio-frequency
(RF) chains increases energy consumption,
and significant latency is introduced
due to the precoding processing.
These physical limitations have 
recently prompted a shift towards a disruptive 
new technology, renewing interest
in previously abandoned analog computing approaches.

Capitalizing on the isomorphism between
the Huygens-Fresnel principle and the architecture of a dense neural network, 
a free-space all-optical 
{\em diffractive deep neural network ($\text{D}^2$NN)}
was developed in \cite{Lin.2018}. This network 
performs computations or inference tasks via wave propagation
and diffraction through a series of stacked metasurfaces. 
Essentially, a $\text{D}^2$NN processes the wavefront of the input 
electromagnetic (EM) field as 
it propagates through a series of structured diffractive metasurfaces 
connected via free-space propagation, with each meta-atom functioning
as an artificial neuron. 

Diffractive optical processing holds the potential for massively
parallel computations, enabling light-speed calculations with low power consumption. It also provides  
direct access to all fundamental properties of input waves, including amplitude, phase, 
polarization, and orbital angular momentum. 
Inspired by the pioneering work in \cite{Lin.2018}, 
recent years have seen growing interest in free-space optical computing platforms for applications in statistical inference, computational
imaging, and sensing \cite{Liu.2022,Sun.2023}. 

In wireless communications, the fundamental physical principles of 
diffractive optical platforms have inspired the development of 
{\em stacked intelligent metasurfaces (SIM)} \cite{Hassan.2024, 
Nerini.2024,DiRenzo, Hanzo, Di Renzo-ICC,Liu.2024,Yao.2024,Lin.2024,Liu-arXiv_2024}, which 
consist of cascades of programmable metasurfaces, each with adaptive or intelligent capabilities.
SIM enable the implementation of signal-processing transformations directly 
in the EM wave domain, since they 
produce a wave profile at the output by suitably tailoring 
the input wave as it propagates through them.
The integration of SIM into wireless communication systems 
enables the replacement of conventional digital beamforming
structures. This shift  reduces the need for high-resolution DACs/ADCs and decreases the number of RF chains, leading to lower
hardware costs and power consumption. 
Additionally, transmit precoding and receiver
combining occur in the wave domain as the  
EM signal
propagates through the SIM  at light speed, thereby reducing 
processing delay compared to digital systems.
Furthermore, SIM expand the range of
two-dimensional (2-D) complex-valued transformations that can 
be applied to the input RF signal \cite{Hassan.2024, 
Nerini.2024,DiRenzo, Hanzo, Di Renzo-ICC,Liu.2024,Yao.2024,Lin.2024,Liu-arXiv_2024}.

\subsection{Related works}

In \cite{Hassan.2024}, a free-space path-loss model was developed 
to compute the received signal power at a 
single-antenna receiver when the transmitter is equipped 
with a SIM module. This model was also  used 
to maximize the power at the receiver. 
Additionally,  \cite{Nerini.2024} presents a comprehensive
model of a SIM-aided communication system using
multiport network theory. This model accounts for mutual coupling 
effects at the transmitter, SIM layers, and receiver.
In \cite{DiRenzo}, the authors proposed using SIM for 2-D direction-of-arrival estimation
of a single source transmitter. By suitably configuring the
phase shifts of the SIM, the receiver antenna array 
can directly observe the angular spectrum of the incident signal.
Similarly, \cite{Hanzo} describes a holographic MIMO communication system equipped 
with SIM integrated at both the transmitter and receiver. In this system,
the end-to-end channel matrix is transformed into
non-interfering parallel subchannels though 
joint optimization of the phase shifts across 
metasurface layers of both the transmit and receive SIM.
The above studies \cite{Hassan.2024,Nerini.2024,DiRenzo,Hanzo}
focus on {\em point-to-point SIM-aided communication}, which involves scenarios with
a single transmitter and a single receiver.

The application of SIM to {\em multiuser beamforming} in a multiple-input 
single-output (MISO) downlink system was explored in 
\cite{Di Renzo-ICC,Liu.2024}.
Specifically, \cite{Di Renzo-ICC} focuses on maximizing
the sum rate of all users  by jointly optimizing the transmit
power allocation and the wave-based 
precoding implemented by the SIM at the base station (BS).
This results in a non-convex design problem, which is addressed using
alternating optimization (AO).
In \cite{Liu.2024},
a solution to this non-convex optimization problem is proposed,  employing deep reinforcement learning (DRL).

Maximizing the sum rate in a downlink system requires knowledge of 
instantaneous channel state information (CSI) at the BS. 
The challenge of acquiring instantaneous CSI in SIM-assisted multiuser downlink systems was addressed in \cite{Yao.2024}. 
To address the difficulties of obtaining instantaneous CSI
at the BS, \cite{Lin.2024} formulates
a joint power allocation and SIM phase
shift optimization problem based on statistical CSI to maximize the sum rate.

The works \cite{Hassan.2024, 
Nerini.2024,DiRenzo, Hanzo, Di Renzo-ICC,Liu.2024,Yao.2024,Lin.2024}
focus {\em exclusively} on phase reconfiguration of the SIM. 
Although phase-only SIM are nearly passive and
can be implemented using simple components, such as
tunable varactor diodes or switchable positive-intrinsic-negative
(PIN) diodes, phase control alone does not compensate for propagation losses inside the SIM structure.
These losses can become significant, especially with a large number of layers. 
In principle, one may design the layers of the SIM to jointly control the phase 
and amplitude of the EM transmission coefficients of the meta-atoms
while allowing for amplification. However, to the best of our knowledge, 
simultaneous and independent control of amplitude and phase
for metasurfaces has been experimentally demonstrated only
in the case of attenuation, i.e., 
the magnitude of the EM transmission/reflection coefficients is smaller than or 
equal to one (see \cite{Liao_2022} for a single-layer metasurface 
working  in reflection mode).
In this paper, capitalizing on the multilayer structure of SIM, we propose
to combine phase-only and amplitude-only layers to achieve 
the best balance between manipulation capabilities and system complexity.

\begin{figure*}[t]
\centering
\includegraphics[width=\columnwidth]{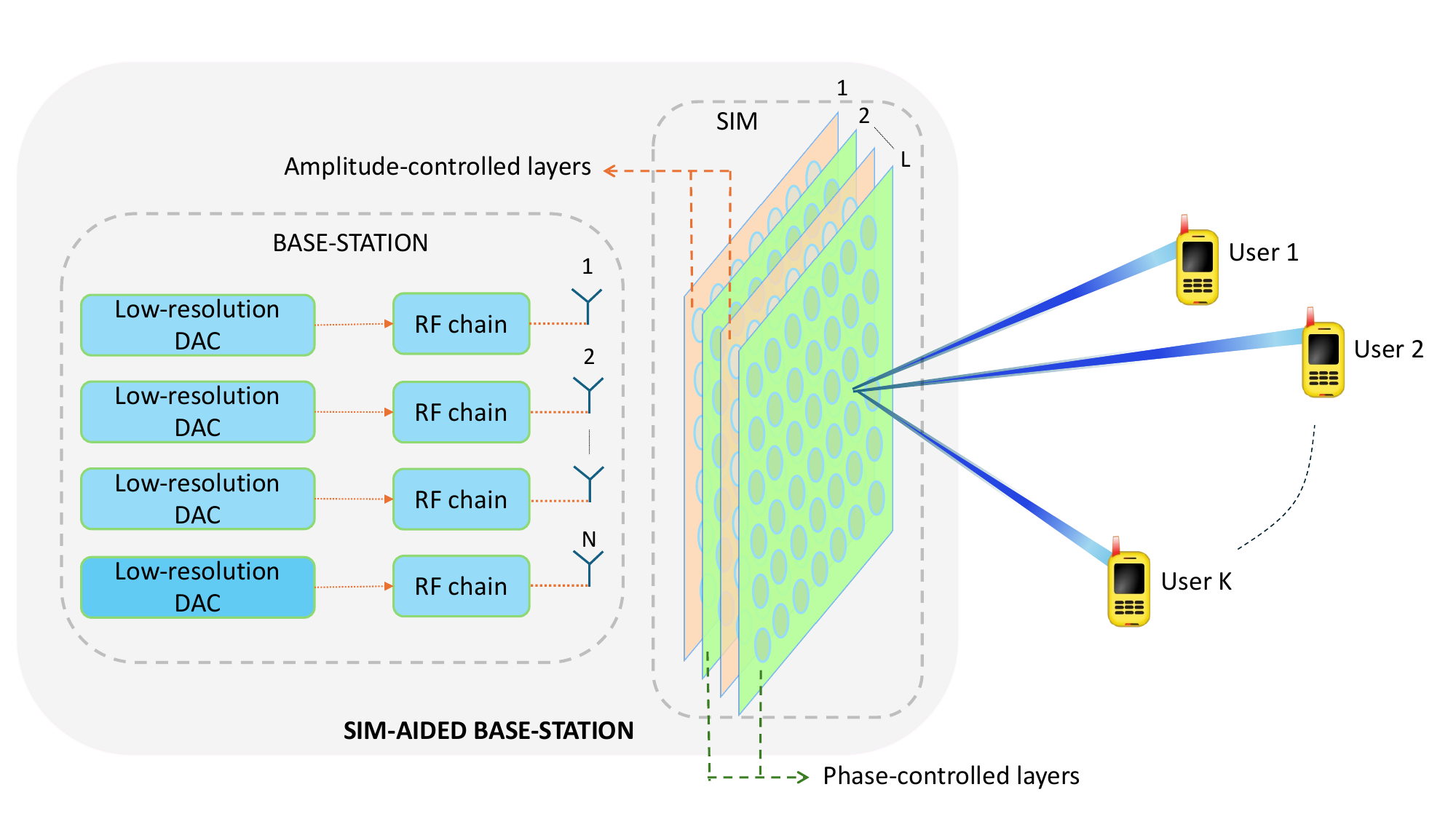}
\caption{Schematic of SIM-aided multiuser downlink transmission, with the AP and AC layers alternated
for the sake of simplicity.}
\label{fig:fig_1}
\end{figure*}

\subsection{Contribution and organization}

Recently,  \cite{Liu.2022} demonstrated a $\text{D}^2$NN capable of hierarchically manipulating the energy distribution of transmitted 
EM waves using a stack of metasurfaces. In this implementation,  
the amplitude of the transmitted wave through each meta-atom 
is adjusted by controlling
amplifier chips via field programmable gate arrays
(FPGAs).

Building on this experimental platform, we propose
a novel SIM configuration in Section~\ref{sec:model} that incorporates
both {\em phase-controlled (PC)}  
and {\em amplitude-controlled (AC)} layers.
Specifically, PC layers are nearly passive, allowing only phase adjustments of
their meta-atoms via components such as 
varactor or PIN diodes. In contrast, AC layers are active 
and enable amplitude modulation of their meta-atoms through the integration of amplifier chips. This combination  
of PC and AC layers in the SIM facilitates independent manipulation 
of amplitude and phase in the wave domain.

Compared to PC-only SIM 
\cite{Di Renzo-ICC,Liu.2024}, the inclusion of AC layers provides additional degrees 
of freedom that can enhance downlink beamforming design (see
Section~\ref{sec:sched-bf-opt}). In addition to 
the standard {\em sum-power constraint} at the BS, our approach incorporates
a {\em per-stream power preserving constraint} to mitigate propagation losses across the SIM and 
an {\em amplitude constraint}  
to account for power limitations for
each active layer. In this framework, 
the key contributions of this study are
summarized as follows.

\begin{enumerate}

\item
In Section~\ref{sec:opt-bf}, 
we address  the {\em optimal} beamforming design aimed at 
maximizing the sum rate of a group of users. This is achieved
by jointly optimizing the 
transmit power allocation at the BS and
the wave-based beamforming at the SIM, while adhering to both 
sum-power and peak-power constraints.
To enhance system throughput, the BS 
{\em opportunistically} schedules transmission to users with favorable channel
fading conditions, leveraging {\em multiuser diversity} effects \cite{Tse-book}.
The problem is approached in two stages:
first, we synthesize the optimal beamforming matrix, 
which includes the transmission coefficients of the 
meta-atoms and the Rayleigh-Sommerfeld diffraction parameters.
In the second stage, we determine the optimal transmission
coefficients by finding the 
best fit in a least-squares sense 
between the designed beamforming matrix and 
the EM response of the SIM.

\item
In Section~\ref{sec:zfbf}, we develop a low-complexity yet suboptimal 
{\em zero-forcing (ZF) beamforming} scheme, which enforces the SIM 
to eliminate interference among user streams. By combining this strategy 
with opportunistic user selection, the scheme demonstrates fairly good 
performance when the number of meta-atoms is sufficiently larger than
the number of users.

\item 
In Section~\ref{sec:numer},
Monte Carlo simulations highlight the advantages of incorporating 
AC layers in the SIM by assessing the 
channel capacity under various configurations. Additionally,
we demonstrate that, in the case of favorable propagation,  
the sum-rate  performance of the ZF beamforming combined with opportunistic 
user selection approaches optimality.

\end{enumerate}

These key results are also summarized in Section~\ref{sec:concl},
along with potential pathways for advancing SIM-based wireless communications.

\subsection{Main notations}
\label{sec:pre}

Upper- and lower-case bold letters denote matrices and vectors;
the superscripts
$*$, $\trasp$ and $\herm$
denote the conjugate,
transpose, and Hermitian (conjugate transpose) of a matrix;
$\mathbb{C}$, $\mathbb{R}$ and $\mathbb{Z}$ are
the fields of complex, real and integer numbers;
$\mathbb{C}^{n}$ $[\mathbb{R}^{n}]$ denotes the
vector-space of all $n$-column vectors with complex
[real] coordinates;
similarly, $\mathbb{C}^{n \times m}$ $[\mathbb{R}^{n \times m}]$
denotes the vector-space of all the $n \times m$ matrices with
complex [real] elements;
$j \eqdef \sqrt{-1}$ denotes the imaginary unit;
$\Re(x)$ is the real part of $x \in \Cset$;
$\Im(x)$ is the imaginary part of $x \in \Cset$;
$\delta(\cdot)$ denotes the Dirac distribution; 
$[x]^{+}$ stands for $\max\{x,0\}$;
$\nabla_{\bm x}[f(\bm x)]$ is the gradient of the function $f(\bm x)$;
$\nabla^2 = \frac{\partial^2}{\partial x^2}+\frac{\partial^2}{\partial y^2}+\frac{\partial^2}{\partial z^2}$
is the three-dimensional Laplace operator;
$\psi(t)$ is the root raised-cosine filter having $\psi(0) = 1$;
$\circ$ is the Hadamard product; 
matrix $\mathbf{A}= \diag (a_{0}, a_{1}, \ldots,
a_{n-1}) \in \mathbb{C}^{n \times n}$ is diagonal;
$\mathbf{1}_n \eqdef [1, \ldots, 1]^T$ is the all-ones vector;
$\{\mathbf{A}\}_{i,\ell}$ indicates the
$(i,\ell)$-th element of $\mathbf{A} \in \Cset^{n \times m}$, with 
$i \in \{1,2,\ldots, n\}$ and $\ell \in \{1,2,\ldots, m\}$;
$\{\mathbf{a}\}_{\ell}$ indicates the
$\ell$-th element of $\mathbf{a} \in \Cset^{n \times 1}$, with 
$\ell \in \{1,2,\ldots, n\}$;
$\|\mathbf{a}\| = (\bm a^\herm \bm a)^{1/2}$ denotes the 
norm of $\mathbf{a} \in \Cset^{n}$, whereas 
$\|\bm A \| \eqdef [\text{trace}(\bm A \, \bm A^\herm)]^{1/2}$
denotes the Frobenius matrix norm of $\bm A \in \mathbb{C}^{m \times n}$;
$\bm A^{-1}$ and 
$\bm A^{\dag}$ are the inverse and the Moore-Penrose inverse (pseudoinverse) of the matrix $\bm A$, respectively;
$\text{card}(\mathcal{S})$ is the cardinality of the set $\mathcal{S}$;
the Landau notation $\mathcal{O}(n)$ stands for 
``growth at the order of $n$''; 
$\Es[\cdot]$ denotes ensemble averaging.

\section{System model}
\label{sec:model}

We consider the SIM-aided multiuser downlink communication 
system illustrated in Fig.~\ref{fig:fig_1}. In this setup,  
the BS is equipped with $N$ antennas and uses SIM 
consisting of $L$ planar metasurface layers to communicate with $K \ge N$ single-antenna users.
Let $s$ denote the spacing between two adjacent layers of the SIM, and $\sigma$
the distance between the array and the first layer of the SIM.  
Each layer of the SIM is composed of $Q \eqdef Q_{x} \times Q_{y}$ meta-atoms arranged in a rectangular grid with $Q_{x}$ and $Q_{y}$ elements along the $x$ and $y$ axes, respectively, and a constant inter-element spacing
$d_{\text{RIS}}$.
For simplicity, 
we introduce a mapping that converts the 2-D index
$(q_x,q_y)$ of a meta-atom in each layer, where $q_x \in 
\{0, 1, \ldots, Q_{x}-1\}$ and $q_y \in \{0, 1, \ldots, Q_{y}-1\}$, into a one-dimensional (1-D) index 
$q \eqdef q_x \, Q_{y} + q_y$ belonging to $\mathcal{Q} \eqdef \{0, 1, \ldots, Q-1\}$. This mapping indexes 
the meta-atoms sequentially from row to row within each layer.

\subsection{Forward propagation mechanism through the SIM}
\label{sec:prop_SIM}
Wave propagation through stacked metasurfaces is a complex process that generally involves multiple interactions. These interactions can be rigorously analyzed using full-wave numerical simulations (e.g., finite-element methods) or semi-analytical techniques \cite{Menzel.2016}. However, such models are not well-suited for application in communication scenarios.

In this work, we adopt a standard analytical propagation model, originally developed in the context of $\text{D}^2$NNs \cite{Lin.2018, Liu.2022}, and later applied in various SIM studies \cite{Hassan.2024, Nerini.2024, DiRenzo, Hanzo, Di Renzo-ICC, Liu.2024, Yao.2024, Lin.2024}. In SIM scenarios, it is typically assumed that each metasurface layer is perfectly impedance-matched, eliminating reflections and focusing solely on forward propagation. Although this assumption imposes limitations on the operating bandwidth, it allows for an analytical approach that is compatible with communication scenarios and related optimization tools.
When an incident EM wave with carrier frequency $f_0>0$ passes through a generic meta-atom of the first  metasurface layer, the amplitude and phase of the transmitted wave
are determined by the product of the incident electric field and the complex-valued transmission coefficient. This transmitted wave then acts 
as a secondary source, illuminating all the
meta-atoms in the second metasurface layer, as described by the Huygens-Fresnel principle\cite{Goodman}. 
\begin{table*}
\caption{Main system parameters.} 
\label{tab:tab-1}
\centering{}%
\begin{tabular}{cc}
\hline
\noalign{\vskip\doublerulesep}
\textbf{Symbol} & \textbf{Meaning} 
\tabularnewline[\doublerulesep]
\hline
$N$ & Number of transmit antennas 
\tabularnewline
$K$ & Number of system users 
\tabularnewline
$L$ & Number of metasurface layers
\tabularnewline
$s$ & Spacing between adjacent layers of the SIM
\tabularnewline
$\sigma$ & Spacing between the BS array and the first layer of the SIM
\tabularnewline
$d_{\text{RIS}}$ & Spacing between adjacent meta-atoms
\tabularnewline
$Q$ & Number of meta-atoms of the SIM per layer
\tabularnewline
${L}_{\text{pc}}$ & Number of PC layers
\tabularnewline
${L}_{\text{ac}}$ & Number of AC layers
\tabularnewline
$b$ & Number of coding bits for the phase values of PC layers
\tabularnewline
$M=2^b$ & Number of possible phase values for PC layers 
\tabularnewline
\hline
\end{tabular}
\end{table*}
To enhance the degrees of freedom in the wave-domain transformation
performed by the SIM, compared to recent designs
\cite{Hassan.2024, 
Nerini.2024,DiRenzo, Hanzo, Di Renzo-ICC,Liu.2024,Yao.2024,Lin.2024},
we propose incorporating both AC and PC layers into the stacked device (see Fig.~\ref{fig:fig_1}). 
Let $\gamma_{\ell,q} = \alpha_{\ell,q} \,  e^{j \phi_{\ell,q}}$ represent the EM transmission coefficient of the $q$-th meta-atom in the 
$\ell$-th metasurface layer (evaluated at frequency $f_0$), with $q \in \mathcal{Q}$ and 
$\ell \in \mathcal{L} \eqdef \{1, 2, \ldots, L\}$. We define $\mathcal{L}_{\text{ac}}$ and 
$\mathcal{L}_{\text{pc}}$ as two nonoverlapping subsets of $\mathcal{L}$ 
that index the AC and PC layers, respectively, whose 
cardinalities $L_{\text{ac}}$ and $L_{\text{pc}}$ satisfy the condition 
$L_{\text{ac}} + L_{\text{pc}}=L$.\footnote{Mathematically, the subsets $\mathcal{L}_{\text{ac}}$ and
$\mathcal{L}_{\text{pc}}$ determine a partition of the set $\mathcal{L}$, i.e., 
$\mathcal{L}_{\text{ac}} \cup \mathcal{L}_{\text{pc}}=\mathcal{L}$
and $\mathcal{L}_{\text{ac}} \cap \mathcal{L}_{\text{pc}}=\emptyset$.
} 
The transmission coefficients for each layer of the SIM are organized into diagonal matrices 
$\bm \Gamma_{\ell} \eqdef \diag(\pmb \gamma_\ell) \in \Cset^{Q \times Q}$, where
$\pmb \gamma_\ell \eqdef [\gamma_{\ell,0}, \gamma_{\ell,1}, \ldots, \gamma_{\ell,Q-1}]^\trasp$ and $\ell \in \mathcal{L}$.

AC layers consist of meta-atoms with amplitude responses 
$\{\alpha_{\ell,q}\}_{\ell \in \mathcal{L}_{\text{ac}}}$ that
can be independently controlled through software. This amplitude control is achieved by 
integrating amplifier chips in each meta-atom of the AC layers \cite{Liu.2022}, allowing for a substantial dynamic modulation range.\footnote{In \cite{Liu.2022}, 
each meta-atom is equipped with two amplifier chips, providing
a dynamic modulation range of $35$ dB.
}
The phases $\{\phi_{\ell,q}\}_{\ell \in \mathcal{L}_{\text{ac}}}$  of the transmission coefficients 
in AC layers are fixed and cannot be adjusted. Consequently, these phases will be treated as known but uncontrollable in the subsequent 
optimization process.

For the PC layers, the metasurfaces are locally passive, i.e., 
their meta-atoms cannot amplify the incident EM waves. 
Due to the inevitable material losses, PC layers 
may atte\-nua\-te the EM waves
that penetrate through them, implying that 
their amplitude responses are in general smaller than or equal to one, 
i.e., $\alpha_{\ell,q} \le 1$ for $\ell \in \mathcal{L}_{\text{pc}}$.
However, different from the propagation losses between adjacent layers, 
transmission losses can be controlled and reduced by appropriate designs \cite{Bing.2020}. 
Therefore, we assume that the PC layers have a constant 
transmittance, meaning that the amplitude responses are fixed at $\alpha_{\ell,q}=\alpha_\text{pc} \le 1$ 
for $\ell \in \mathcal{L}_{\text{pc}}$. 
The phases $\{\phi_{\ell,q}\}_{\ell \in \mathcal{L}_{\text{pc}}}$ can be adjusted within the interval $[0,2\pi)$.  
Specifically, for an $b$-bit
digital meta-atom in a PC layer, 
each phase $\phi_{\ell,q}$ can take on values from a 
set 
$\Phi \eqdef \{\phi_0, \phi_1, \ldots, \phi_{M-1}\}$
with cardinality $M \eqdef 2^b$. These transmission phases are given 
by
$\phi_m \eqdef e^{j \frac{2 \pi}{M} m}$, for $m \in \mathcal{M} \eqdef \{0,1, \ldots, M-1\}$.
The channel coefficients between the $N$ 
transmit antennas of the BS array
and the $Q$ meta-atoms of the first layer of the SIM are organized into the 
matrix $\bm W_1 \in \Cset^{Q \times N}$. These coefficients
are modeled using  {\em Rayleigh-Sommerfeld diffraction 
theory} as follows (see Appendix~\ref{sec:appendix} for details)
\be
\{\bm W_1\}_{q,n} = \frac{A_{\text{bs}} \, \cos(\theta^{(1)}_{q,n})}{d^{(1)}_{q,n}}
\left(\frac{1}{2 \pi d^{(1)}_{q,n}}-\frac{j}{\lambda_0}\right) \, 
e^{j \frac{2 \pi}{\lambda_0} {d^{(1)}_{q,n}}}
\label{eq:W1}
\ee
for $q \in \mathcal{Q}$ and $n \in \mathcal{N} \eqdef \{0, 1, \ldots, N-1\}$, 
where $\lambda_0 = c/f_0$ is the wavelength, 
with $c = 3 \cdot 10^8$ m/s denoting the light speed in  vacuum, 
$A_{\text{bs}}$ is the effective area of the antennas of the array (evaluated at $f_0$), 
$\cos(\theta^{(1)}_{q,n})=\sigma/d^{(1)}_{q,n}$, and $d^{(1)}_{q,n}$ denotes the distance between the $n$-th antenna of the BS and the $q$-th meta-atom of the first layer. 
This distance reads as
\be
d^{(1)}_{q,n} =  \sqrt{\left(x_q^{(1)}-x_n^{(0)}\right)^2 + \left(y_q^{(1)}-y_n^{(0)}\right)^2 +\sigma^2}
\label{eq:dqq-1}
\ee
with  $(x_q^{(1)}, y_q^{(1)})$ and  $(x_n^{(0)}, y_n^{(0)})$ 
representing the $2$-D position of the $q$-th meta-atom on the first layer of the SIM
and the position of the $n$-th antenna of the BS, respectively.\footnote{Conventionally, the layer $\ell=0$ is considered to be at the position of the BS the antenna array.}

Similarly, for any $\ell \in \mathcal{L}-\{1\}$, 
the forward propagation process between layers $\ell-1$ and $\ell$ is 
described by the matrix 
$\bm W_\ell \in \mathbb{C}^{Q \times Q}$, whose 
entry $\{\bm W_\ell\}_{q,\tilde{q}}$ represents
the channel coefficient from the $\tilde{q}$-th meta-atom in the $(\ell-1)$-th layer 
to the $q$-th meta-atom in the $\ell$-th layer, and is given by 
(see Appendix~\ref{sec:appendix} again)
\be
\{\bm W_\ell\}_{q,\tilde{q}} = \frac{A_{\text{meta}} \, \cos(\theta^{(\ell)}_{q,\tilde{q}})}{d^{(\ell)}_{q,\tilde{q}}} 
\left(\frac{1}{2 \pi d^{(\ell)}_{q,\tilde{q}}}-\frac{j}{\lambda_0}\right) \, 
e^{j \frac{2 \pi}{\lambda_0} {d^{(\ell)}_{q,\tilde{q}}}}
\label{eq:Well}
\ee
where $A_{\text{meta}}$ is the physical area of the meta-atoms, 
$\cos(\theta^{(\ell)}_{q,\tilde{q}})={s}/{d^{(\ell)}_{q,\tilde{q}}}$, and $d^{(\ell)}_{q,\tilde{q}}$ represents the propagation distance between 
the $\tilde{q}$-th meta-atom of the $(\ell-1)$-th layer
and the ${q}$-th meta-atom of the $\ell$-th layer. This distance is given by the following expression
\be
d^{(\ell)}_{q,\tilde{q}} =  \sqrt{\left(x_q^{(\ell)}-x_{\tilde{q}}^{(\ell-1)}\right)^2 + 
\left(y_q^{(\ell)}-y_{\tilde{q}}^{(\ell-1)}\right)^2 + s^2} 
\label{eq:dqq}
\ee
with  $(x_q^{(\ell)}, y_q^{(\ell)})$ representing the $2$-D position of the $q$-th meta-atom 
on the $\ell$-th layer of the SIM.

The overall forward propagation through the SIM can be described by the matrix
$\bm G \in \Cset^{Q \times N}$ given by 
\be
\bm G =  \bm \Gamma_{L} \, \bm W_{L} \, \bm \Gamma_{L-1} \, 
\bm W_{L-1} \cdots \bm \Gamma_2 \bm W_2 \, \bm \Gamma_1 \, 
\bm W_1 \: .
\label{eq:forward}
\ee
The key system parameters are summarized in Table~\ref{tab:tab-1}.

\subsection{Signal radiated by the last layer of the SIM}

For $i \in \mathcal{N}$, 
the complex envelope of the narrowband continuous-time 
signal asso\-cia\-ted with the $i$-th data stream is given by
\be
x_i(t) = \sqrt{\euscr{P}_i} \sum_{n=-\infty}^{+\infty}  b_i(n) \, \psi(t-n \, T_\text s)
\ee
where $\euscr{P}_i$ denotes the transmit power for the $i$-th stream, and
$b_0(n), b_1(n), \ldots, b_{N-1}(n)$ are mutually independent sequences of zero-mean
unit-variance independent and identically-distributed (i.i.d.) 
circularly-symmetric complex random variables. These sequences are transmitted at a 
baud rate $1/T_\text s$, 
and $\psi(t)$ represents the unit-energy square-root Nyquist pulse-shaping
filter.

Let $\bm x(t) \eqdef [x_0(t), x_1(t), \ldots, x_{N-1}(t)]^\trasp \in \Cset^N$ represent the complex envelope of the 
signal transmitted by the antenna array at the BS. The baseband signal  sent  from the $L$-th layer of the SIM and propagating through the physical channel is given by
\be
\bm z(t) = \bm G \, \bm x(t) = \sum_{i=0}^{N-1} \bm g_i \, x_i(t)
\label{eq:z}
\ee
where $\bm g_i \in \Cset^{Q}$ is the $i$-th column of the beamforming matrix $\bm G =  [\bm g_0, \bm g_1, \ldots, \bm g_{N-1}]$.
Using \eqref{eq:z}, the total power radiated by the SIM is given by
\begin{eqnarray}
\euscr{P}_{\text{rad}} \eqdef \Es[\|\bm z(t)\|^2] 
&=& \sum_{i_1=0}^{N-1} \sum_{i_2=0}^{N-1} \bm g^\herm_{i_1} \, \bm g_{i_2} \,
\Es  \left[ x_{i_1}(t) \, x^*_{i_2}(t)\right ]  \nonumber \\
&=& \sum_{i=0}^{N-1} \euscr{P}_{i} \, \|\bm g_{i}\|^2 
\label{eq:Prad}
\end{eqnarray}
where we have exploited the statistical independence among the information symbols $b_i(n)$.

\subsection{Signal received by the users}
\label{sec:model-user}

For $k \in \mathcal{K} \eqdef \{1, 2, \ldots, K\}$, at the receiver of the $k$-th user, and under 
the assumption of perfect time and frequency synchronization, 
the discrete-time signal
after down-conversion, matched 
filtering, and sampling with rate $1/T_\text s$,  is given by
\begin{eqnarray}
y_k(n) = \sqrt{\varrho_k} \, \bm h_k^H  \, \sum_{i=0}^{N-1} \sqrt{\euscr{P}_i} \,   \bm g_i \, b_i(n) + w_k(n)
\label{eq:yk}
\end{eqnarray}
where the low-pass equivalent response $\bm h_k \in \Cset^{Q}$ models the frequency-flat block fading channel from the 
$L$-th layer of the SIM to the $k$-th user, with $\Es[\|\bm h_k\|^2]=Q$, the scalar factor
$\varrho_k \eqdef [\lambda^2/(4 \, \pi \, d_0)^2] \, (d_0/d_k)^\eta$ accounts for the propagation path-loss of the $k$-th user link, $d_k$ is the 
distance between the SIM and the $k$-th user, $d_0$ is a reference distance for the antenna far field, and $\eta$ is the path-loss exponent.
The noise $w_k(n)$ is modeled as a sequence of i.i.d.
circularly-symmetric complex Gaussian random variables, with zero mean and variance 
$\sigma_{w}^2$, statistically independent of $b_i(n)$ for any $i \in \mathcal{N}$ and $n \in \mathbb{Z}$.

\section{Joint scheduling and beamforming optimization}
\label{sec:sched-bf-opt}

User streams are separated by different beamforming
directions, which are determined by the columns of the matrix $\bm G$. This matrix, in turn, 
depends on the transmission coefficients of the SIM as described by  
\eqref{eq:forward}. 
In this context, we assume the presence of a multiuser scheduler at the BS, 
whose aim is to select a subset $\mathcal{K}_N  \eqdef \{k_0, k_1, \ldots, k_{N-1}\} \subset \mathcal{K}$ 
of the network users, with $\text{card}(\mathcal{K}_N) = N$,
to maximize the overall system throughput.
Our optimization variables include the scheduling subset $\mathcal{K}_N$, the 
diagonal entries of the matrices $\bm \Gamma_{1}, \bm \Gamma_{2}, \ldots, \bm \Gamma_{L}$, and the transmission powers allocated to the data streams for the scheduled users.
The rate allocation is based on the full CSI available at the BS.
For details on acquiring this CSI  in SIM-assisted multiuser downlink systems, the reader is referred to \cite{Yao.2024}.

Let $\bm H = [\bm h_1^*, \ldots, \bm h_K^*]^\trasp \in \mathbb{C}^{K \times Q}$ denote the matrix of user channel vectors. We define the {\em reduced} channel matrix $\widetilde{\bm H}(\mathcal{K}_N)$ by selecting
the rows of $\bm H$ corresponding to the indices in the subset $\mathcal{K}_N$:
\be
\widetilde{\bm H}(\mathcal{K}_N) = [\bm h_{k_0}^*, \bm h_{k_1}^*, \ldots, \bm h_{k_{N-1}}^*]^\trasp
\in \mathbb{C}^{N \times Q} \:. 
\label{eq:Htilde}
\ee
For notation simplicity, we map the entries of $\mathcal{K}_N$ to a set of natural numbers such that
$k_i \rightarrow i$, where $i \in \mathcal{N}$. Accordingly, we rename the ($k_i$)-th row of ${\bm H}$ as
$ \widetilde{\bm h}_{i}$, thus obtaining
$\widetilde{\bm H}(\mathcal{K}_N) = [\widetilde{\bm h}_{0}^*, \widetilde{\bm h}_{1}^*, \ldots, \widetilde{\bm h}_{N-1}^*]^\trasp$.
The propagation constants corresponding to the channel matrix 
$\widetilde{\bm H}(\mathcal{K}_N)$ are denoted with $\widetilde{\varrho}_i$, for 
$i \in \mathcal{N}$.
Similarly, we indicate with 
\be
\widetilde{\bm G}(\mathcal{K}_N) = [\widetilde{\bm g}_{0}, \widetilde{\bm g}_{1}, \ldots, \widetilde{\bm g}_{N-1}] \in \mathbb{C}^{Q \times N} 
\label{eq:GKN}
\ee
the  matrix collecting the beamforming vectors of the scheduled users, whereas 
$\widetilde{\euscr{P}}_0(\mathcal{K}_N), \widetilde{\euscr{P}}_1(\mathcal{K}_N), \ldots, 
\widetilde{\euscr{P}}_{N-1}(\mathcal{K}_N)$ are the corresponding transmit powers.
The {\em signal-to-interference-plus-noise ratio (SINR)} at the $i$-th scheduled user can be expressed as
\be
\text{SINR}_i(\mathcal{K}_N) = \frac{\widetilde{\varrho}_i \, \widetilde{\euscr{P}}_i(\mathcal{K}_N) \,  |\widetilde{\bm h}_i^H  \, \widetilde{\bm g}_i |^2}
{\displaystyle  \widetilde{\varrho}_{i} \sum_{\shortstack{\footnotesize $i'=0$ \\ \footnotesize $i' \neq i$}}^{N-1}  \widetilde{\euscr{P}}_{i'}(\mathcal{K}_N) 
\, |\widetilde{\bm h}_i^H  \, \widetilde{\bm g}_{i'}|^2 + \sigma_w^2} \:.
\label{eq:SINR}
\ee
Assuming that the transmitter encodes the information 
of each user by using an i.i.d. Gaussian code, 
the {\em sum-rate capacity} for the user group $\mathcal{K}_N$ is given by 
\be
\rate(\mathcal{K}_N) = \sum_{i =0}^{N-1} \log_2\left[1+\text{SINR}_i(\mathcal{K}_N)\right] \quad \text{(bit/s/Hz)} \: .
\label{eq:sum-rate}
\ee

The proposed joint scheduling and beamforming algorithm involves the following three main stages
(in the given order).
\begin{enumerate}
\item 
{\em Beamforming optimization}:
For each subset $\mathcal{K}_N$, we seek to maximize in Subsection~IV-\ref{sec:first-prob} 
the sum-rate capacity \eqref{eq:sum-rate} by adjusting the beamforming directions
and the power allocations for the users in $\mathcal{K}_N$,
under specific constraints that will be defined in Subsection~III-\ref{sec:power-constr}.
The beamforming optimization with the additional ZF constraint
is instead solved in Subsections~V-\ref{sec:step-1-zf} and V-\ref{sec:step-2-zf}.
\item
{\em Scheduling optimization}:
Once the beamforming vectors and power allocations 
for a given $\mathcal{K}_N$ are determined, we find the optimal
 subset $\mathcal{K}^\star_N$ of users by performing an exhaustive search
\be
\mathcal{K}^\star_N \eqdef \arg \max_{\mathcal{K}_N \subset \mathcal{K}} \, \rate(\mathcal{K}_N) 
\label{eq:scheluding-opt}
\ee
which identifies the subset that maximizes the sum capacity. 
To make this search computationally feasible, it should be performed over
relatively small user pools (see the discussion in Subsection~IV-\ref{sec:comp-opt}).
\item
{\em SIM optimization}:
The transmission coefficients of the SIM are obtained starting from the
optimized beamforming matrix. Such an optimization is carried out 
in Subsections~IV-\ref{sec:mod-algo} and IV-\ref{sec:dis-phase}.
\end{enumerate}

To clearly define the beamforming and SIM optimization stages, we need to impose
reasonable constraints that reflect the fundamental physical principles of the SIM and the limited power budget at the BS.
The inclusion of active layers introduces power constraints at the BS that 
are different from those typically associated with 
PC-only SIM  \cite{Di Renzo-ICC,Liu.2024}. This issue is 
addressed in the following subsection.

\subsection{Power constraints at the BS}
\label{sec:power-constr}

The power globally radiated by the SIM-based transmitter is equal to
\eqref{eq:Prad}. To understand its dependence on the SIM parameters, we provide an upper bound on $\euscr{P}_{\text{rad}}$
for a generic beamforming matrix $\bm G$ and power distribution $\{\euscr{P}_{i}\}_{i \in \mathcal{N}}$.

Applying the Rayleigh-Ritz theorem \cite{Horn}, we know that
$\|\bm g_{i}\|^2  \le \beta_\text{max}({\bm G}^\herm \, \bm G)$, where 
$\beta_\text{max}({\bm G}^\herm \, \bm G)$ represents the largest eigenvalue of the
Hermitian matrix ${\bm G}^\herm \, \bm G$, for each $i \in \mathcal{N}$.
Therefore, each $\|\bm g_{i}\|^2$ in \eqref{eq:Prad} is bounded by
the squared spectral norm of $\bm G$ \cite{Horn}. 
According to \eqref{eq:forward}, the matrix ${\bm G}^\herm \, \bm G$ is
the product of positive semi-definite Hermitian matrices. 
By invoking the submultiplicative property of the spectral norm \cite{Horn}, 
we obtain
\barr
\beta_\text{max}({\bm G}^\herm \, \bm G) & \le \prod_{\ell=1}^L 
\beta_\text{max}(\bm W_{\ell}^\herm \, \bm \Gamma_{\ell}^* \, 
\bm \Gamma_{\ell} \, \bm W_{\ell})
\nonumber \\ & \le 
\left[\prod_{\ell=1}^L 
\beta_\text{max}(\bm \Gamma_{\ell}^* \, 
\bm \Gamma_{\ell})\right] 
\left[\prod_{\ell=1}^L 
\beta_\text{max}(\bm W_{\ell}^\herm \, \bm W_{\ell}) \right]
\earr
with 
\be
\beta_\text{max}(\bm \Gamma_{\ell}^* \, 
\bm \Gamma_{\ell}) = \begin{cases}
1\:, & \text{for $\ell \in \mathcal{L}_{\text{pc}}$}    
\\ \max_{q \in \mathcal{Q}} \alpha_{\ell,q}^2\:, &  
\text{for $\ell \in \mathcal{L}_{\text{ac}}$}
\end{cases} \: .
\ee
In practice, the meta-atoms in the AC layers operate within 
a specific amplitude range \cite{Liu.2022}, meaning their 
amplitude responses satisfy
$\alpha_{\ell,q} \in [\alpha_{\text{min}}, \alpha_{\text{max}}]$,
for $\ell \in \mathcal{L}_{\text{ac}}$ and $q \in \mathcal{Q}$.
Consequently, we have
\be
\prod_{\ell=1}^L 
\beta_\text{max}(\bm \Gamma_{\ell}^* \, 
\bm \Gamma_{\ell}) \le \alpha_{\text{max}}^{2 L_{\text{ac}}} \: .
\ee
Using this result, we can derive an upper bound on the radiated power from \eqref{eq:Prad}: 
\be
\euscr{P}_{\text{rad}} \le \alpha_{\text{max}}^{2 L_{\text{ac}}}
\left[\prod_{\ell=1}^L 
\beta_\text{max}(\bm W_{\ell}^\herm \, \bm W_{\ell}) \right]
\left( \sum_{i=0}^{N-1} \euscr{P}_{i} \right) \: .
\label{eq:ineq-pot}
\ee
This expression highlights the influence of three different factors:
the first term represents the impact of the AC layers, showing an exponential 
dependence on their number  $L_{\text{ac}}$; 
the second term accounts for the propagation effects across the SIM;
the third term represents the input power to the SIM.
It is important to note that for 
a PC-only SIM \cite{Di Renzo-ICC,Liu.2024}, where
$L_{\text{ac}}=0$,    
the first term in \eqref{eq:ineq-pot} equals one. This reduces
the available degrees of freedom for controlling $\euscr{P}_{\text{rad}}$.

To prevent propagation losses across the SIM, we enforce the following constraints in our designs:
\be
\|\bm g_i \|^2 = 1 \:, \quad \text{for $i \in \mathcal{N}$}
\label{eq:v}
\ee
which, according to \eqref{eq:Prad},  implies that 
\be
\euscr{P}_{\text{rad}} = \sum_{i=0}^{N-1} \euscr{P}_{i} \: .
\ee
Thus, the power radiated from the SIM exactly matches the input one. 
This power is subject to (s.t.) the standard constraint 
\be
\sum_{i=0}^{N-1} \euscr{P}_{i}  < \euscr{P}_{\text{tot}} 
\label{eq:global}
\ee
where $\euscr{P}_{\text{tot}}>0$ denotes the available power budget at the BS.
We refer to \eqref{eq:v} as the {\em per-stream power preserving constraint}, which 
ensures that the power 
$\euscr{P}_{i}$ allocated to the $i$-th data stream $x_i(t)$ is 
preserved as it travels through the SIM.
On the other hand, the customary requirement \eqref{eq:global}
is referred to as the 
{\em sum-power constraint}.

In \cite{Di Renzo-ICC,Liu.2024}, only the sum-power constraint is considered.
However, it is important to note that 
the per-stream power-preserving constraint can be applied in both 
PC-only SIM or PC-and-AC SIM. For PC-only SIM, according to 
the {\em Huygens-Fresnel principle} \cite{Orfanidis}, the phase
responses of the transmission coefficients of the SIM are
also optimized in such a way the secondary spherical wavelets 
 from the different meta-atoms of the $(\ell-1)$-th layer 
constructively combine on the surface of the 
$\ell$-th layer, for $\ell \in \mathcal{L}-\{1\}$. In the case of 
PC-and-AC SIM, the per-stream power is controlled not only through phase optimization but also through the dynamic adjustment of the amplitude responses 
of the AC layers.

\section{SIM implementing optimal beamforming}
\label{sec:opt-bf}

In this section, we address the {\em optimal} design 
problem of the SIM-based transmitter. When the subset $\mathcal{K}_N$ is fixed, we will use hereinafter the shorthand notations 
$\widetilde{\bm G}$, $\widetilde{\euscr{P}}_i$, 
and $\rate$ in lieu of the more rigorous  ones 
$\widetilde{\bm G}(\mathcal{K}_N)$, $\widetilde{\euscr{P}}_i(\mathcal{K}_N)$,  
and $\rate(\mathcal{K}_N)$, respectively.
The goal is to 
jointly determine the transmission coefficients 
of the SIM and the optimal power allocation for the users 
identified through \eqref{eq:scheluding-opt} that maximize 
the system sum-rate capacity defined by \eqref{eq:SINR} and \eqref{eq:sum-rate}.
Specifically, we aim to optimize the scheduling process, 
the transmission coefficients of the SIM, and the user power allocation 
policy to maximize the objective function
\be
\label{eq:prob-1}
\rate =
\sum_{i =0}^{N-1} \log_2 \left (1  \hspace{-0.5mm} + \frac{\widetilde{\varrho}_i \, \widetilde{\euscr{P}}_i |\widetilde{\bm h}_i^H  \, \widetilde{\bm g}_i|^2}
{\widetilde{\varrho_i} \sum_{\shortstack{\footnotesize $i'=0$ \\ \footnotesize $i' \neq i$}}^{N-1} 
  \widetilde{\euscr{P}}_{i'} |\widetilde{\bm h}_i^H  \widetilde{\bm g}_{i'}|^2 + \sigma_w^2} \right)
\ee
s.t. the following constraints: 
\begin{eqnarray}
\label{eq:constraint}
&& \widetilde{\bm G} = \bm \Gamma_L \, \bm W_{L} \, \bm \Gamma_{L-1} \, \bm W_{L-1} \cdots \bm \Gamma_1 \, \bm W_1 
\label{eq:c1}\\
&& \|\widetilde{\bm g}_i\|^2 = 1\:, \quad i \in \mathcal{N} 
\label{eq:c2-bis} \\
&& \bm \Gamma_\ell = \diag(\pmb{\gamma}_\ell) \:, \quad \ell \in \mathcal{L} 
\label{eq:c3}\\
&& \{\pmb{\gamma}_\ell\}_q = \gamma_{\ell, q} = \begin{cases} \alpha_\text{pc} \, e^{j \, \phi_{\ell, q}} \:, &\ell \in \mathcal{L}_{\text{pc}} 
\label{eq:c4} \\
\alpha_{\ell, q} \, e^{j \, \phi_{\ell, q}} \:, & \ell \in \mathcal{L}_{\text{ac}} \:
\end{cases} 
\label{eq:c4-bis}
\\
&& \alpha_{\text{min}} \le \alpha_{\ell, q} \le \alpha_{\text{max}} \:, \hspace{1.5mm} \ell \in \mathcal{L}_\text{ac} \hspace{1.5mm} \text{and} \hspace{1.5mm} q \in \mathcal{Q} \label{eq:c6} \\
&& \sum_{i=0}^{N-1} \widetilde{\euscr{P}}_{i} \leq \euscr{P}_{\text{tot}} \label{eq:c2}\\
&& \widetilde{\euscr{P}}_i \ge 0 \: , \quad i \in \mathcal{N} \:. 
\label{eq:c5} 
\end{eqnarray}
Constraints \eqref{eq:c1}, \eqref{eq:c3}, and \eqref{eq:c4-bis} account for the specific structure
of the matrices and vectors to be optimized, as detailed in  Subsection~II-\ref{sec:prop_SIM}. 
Eq.~\eqref{eq:c2-bis} enforces the per-stream power preserving constraint, as discussed in
Subsection~III-\ref{sec:power-constr}.
The conventional sum-power constraint is expressed by \eqref{eq:c2} and \eqref{eq:c5}.  
Inequality \eqref{eq:c6} introduces an additional constraint specific to AC layers, referred to as the {\em amplitude constraint}. 
As previously stated in Subsection~III-\ref{sec:power-constr}, this constraint addresses the fact that meta-atoms of AC layers incorporate amplifier
devices. Each meta-atom in these layers acts as a programmable node that modulates the incident 
wave by applying specific voltages to the amplifier chips.\footnote{The modulation is 
assumed to be linear \cite{Liu.2022}.  
While nonlinear modulation could potentially be explored by allowing 
the amplifiers to operate in a nonlinear range, such an approach may compromise system stability. 
} The inherent
relationship between the supply voltage and amplitude modulation imposes
a finite range of the AC transmission coefficients \cite{Liu.2022}.
For the time being, we do not account for the discrete nature of the 
phases $\{\phi_{\ell,q}\}_{\ell \in \mathcal{L}_{\text{pc}}}$ in PC layers. 
We will discuss how such an additional constraint
can be accounted for in Subsection~IV-\ref{sec:dis-phase}.

The constrained optimization problem \eqref{eq:prob-1} is nonconvex, making it challenging to directly determine the 
optimal transmission coefficients and power weights. 
To simplify the process, we decompose the primary optimization problem into three more manageable subproblems. This approach,
known as {\em concentration} in estimation theory literature \cite{Kay.1998}, involves the following steps. 
First, for a given subset $\mathcal{K}_N$, 
we determine the beamforming matrix $\widetilde{\bm G}^\star$ 
and the power allocation policy $\{\widetilde{\euscr{P}}^\star_i\}_{i \in \mathcal{N}}$ that maximize  
the sum-rate, keeping the transmission coefficients of the SIM fixed. 
The optimal subset $\mathcal{K}^\star_N$ of users
is then determined by solving \eqref{eq:scheluding-opt}.
Next, 
based on the previously calculated optimal beamforming matrix 
$\bm G^\star \eqdef \widetilde{\bm G}^\star(\mathcal{K}_N^\star)$,
we compute the optimal transmission coefficients
$\{\pmb{\gamma}^\star_\ell\}_{\ell \in \mathcal{L}}$ characterizing the SIM.

\begin{figure*}[t]
\centering
\includegraphics[width=1.2\columnwidth]{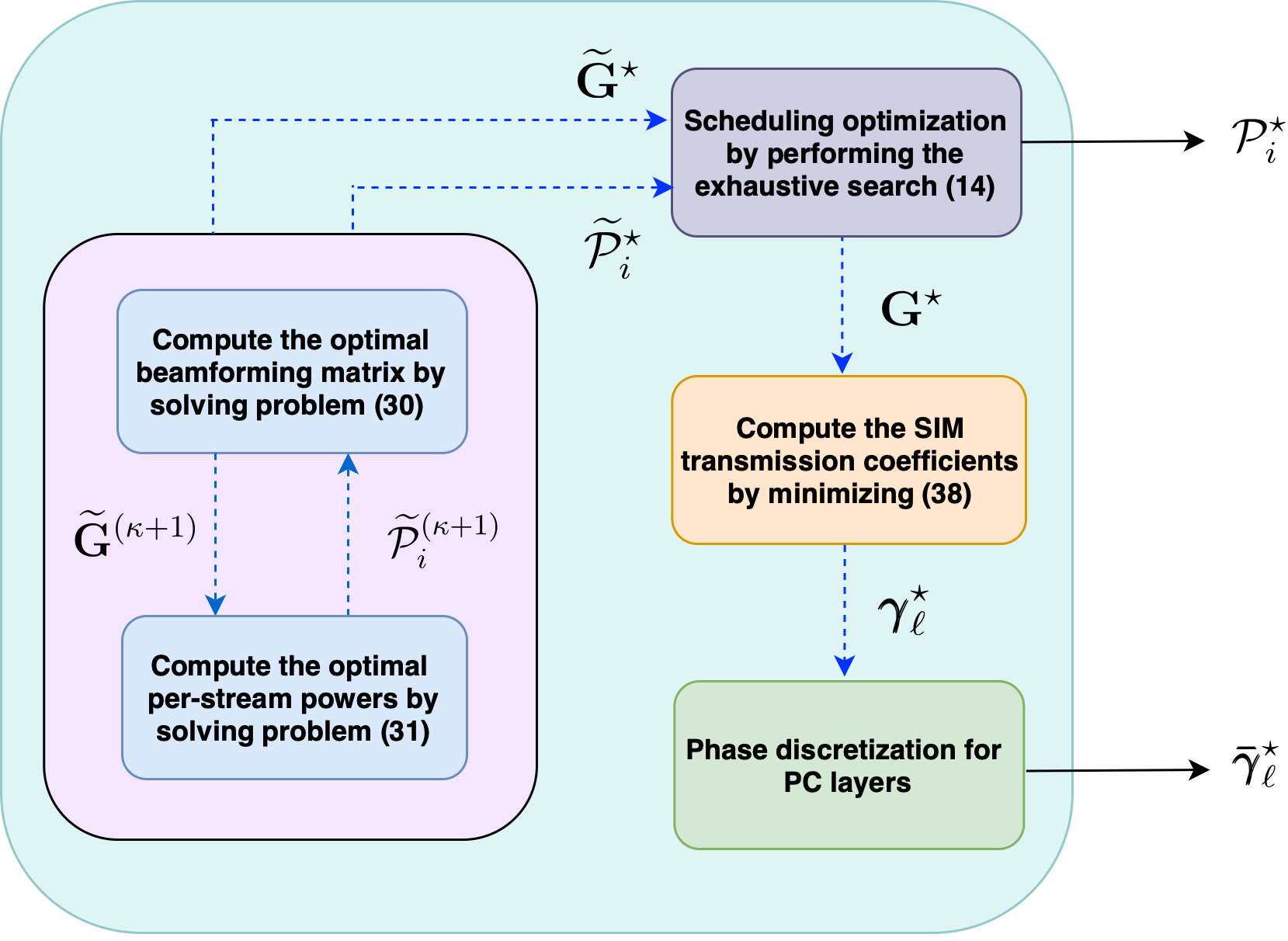}
\caption{Summary of the proposed algorithm maximizing 
the system sum-rate capacity.}
\label{fig:fig_2}
\end{figure*}

The proposed sum-rate capacity maximization procedure is 
summarized in Fig.~\ref{fig:fig_2}.

\subsection{Computation of the beamforming matrix and the per-stream power policy}
\label{sec:first-prob}

The first subproblem involves finding the optimal variables
$(\widetilde{\bm G}^\star,\{\widetilde{\euscr{P}}^\star_i\}_{i \in \mathcal{N}})$ that maximizes
$\rate$ as defined in \eqref{eq:prob-1}, under the constraints
\eqref{eq:c2-bis}, \eqref{eq:c2}, and \eqref{eq:c5}, for a given subset $\mathcal{K}_N$.
To tackle this, we use the {\em block-coordinate
descent (or nonlinear Gauss-Seidel)} method \cite{Bert.1999}, which 
is particularly effective here due to the natural partition of the parameters: 
one set consists of the beamforming matrix $\widetilde{\bm G}$, and the other 
one includes the per-stream powers 
$\{\widetilde{\euscr{P}}_i\}_{i \in \mathcal{N}}$.
Specifically, the next iterates 
$\widetilde{\bm G}^{(\kappa+1)}$ and $\{\widetilde{\euscr{P}}_i^{(\kappa+1)}\}_{i \in \mathcal{N}}$ are generated, given the current 
iterates $\widetilde{\bm G}^{(\kappa)}$ and $\{\widetilde{\euscr{P}}_i^{(\kappa)}\}_{i \in \mathcal{N}}$, according to 
\begin{multline}
\widetilde{\bm G}^{(\kappa+1)}   =
\arg \max_{\text{$\widetilde{\bm G}$ as in \eqref{eq:GKN}}} f_1(\widetilde{\bm G}, \{\widetilde{\euscr{P}}_i^{(\kappa)}\}_{i \in \mathcal{N}}) 
\\ \text{s.t. constraint \eqref{eq:c2-bis}} \label{eq:prob_G-Gauss}
\end{multline}
and
\begin{multline}
\{\widetilde{\euscr{P}}_i^{(\kappa+1)}\}_{i \in \mathcal{N}}  =
\arg \max_{\{\widetilde{\euscr{P}}_i\}_{i \in \mathcal{N}}} f_1(\widetilde{\bm G}^{(\kappa)}, \{\widetilde{\euscr{P}}_i\}_{i \in \mathcal{N}})  
\\ \text{s.t. constraints \eqref{eq:c2} and \eqref{eq:c5}} 
\label{eq:prob_P-Gauss}
\end{multline}
where the cost function $f_1(\widetilde{\bm G}, \{\widetilde{\euscr{P}}_i^{(\kappa)}\}_{i \in \mathcal{N}})$
is obtained from \eqref{eq:prob-1} by replacing 
$\{\widetilde{\euscr{P}}_i\}_{i \in \mathcal{N}}$ with 
$\{\widetilde{\euscr{P}}_i^{(\kappa)}\}_{i \in \mathcal{N}}$ and, similarly, 
$f_1(\widetilde{\bm G}^{(\kappa)}, \{\widetilde{\euscr{P}}_i\}_{i \in \mathcal{N}})$ comes from 
\eqref{eq:prob-1} by taking over $\widetilde{\bm G}$ for $\widetilde{\bm G}^{(\kappa)}$.
It can be shown \cite{Bert.1999} that, if the minimum of the cost functions 
in \eqref{eq:prob_G-Gauss} and \eqref{eq:prob_P-Gauss} is uniquely attained, then every limit point of
$\{\widetilde{\bm G}^{(\kappa)}\}$ and 
$\{\widetilde{\euscr{P}}_0^{(\kappa)}, \widetilde{\euscr{P}}_1^{(\kappa)}, \ldots, \widetilde{\euscr{P}}_{N-1}^{(\kappa)}\}$ is a stationary point
of \eqref{eq:prob-1}, s.t.
\eqref{eq:c2-bis}, \eqref{eq:c2}, and \eqref{eq:c5}.

First, the solution of problem \eqref{eq:prob_G-Gauss} is derived in 
Subsection~IV-A-\ref{sec:step-1}. Then, we find in Subsection~IV-A-\ref{sec:step-2}
the optimal power policy given by \eqref{eq:prob_P-Gauss}.  

\subsubsection{Step $1$: Updating rule of the beamforming matrix}
\label{sec:step-1}

The constrained optimization problem \eqref{eq:prob_G-Gauss} is still nonconvex.
It can be solved using the {\em projected gradient ascent (PGA)} algorithm \cite{Boyd}, which extends the
gradient ascent method to handle constrained maximization problems.
Let 
\be
\Omega \eqdef \left\{\widetilde{\bm g} \in \Cset^Q \,: \,  
\|\widetilde{\bm g}\|^2 = 1 \right\}
\ee
be the feasible set of the columns of $\widetilde{\bm G}$ in 
\eqref{eq:prob_G-Gauss}. Starting from the initial guesses $\widetilde{\bm g}^{(0)}_i$, 
the PGA algorithm is based on the simple iteration
\be
\widetilde{\bm g}^{(\kappa+1)}_i = \mathcal{P}_\Omega \left\{\widetilde{\bm g}^{(\kappa)}_i + \eta^{(\kappa)} \, 
\nabla_{\widetilde{\bm g}_i} f(\widetilde{\bm G}^{(\kappa)}) \right\}
\label{eq:PGA}
\ee
for any $i \in \mathcal{N}$, where $\eta^{(\kappa)}$ is a suitably chosen step length, the function
$f(\widetilde{\bm G}^{(\kappa)})$ is defined in \eqref{eqn_dbl_x}, 
and, for an arbitrary point $\bm g \in \Cset^Q$,  
the operator $\mathcal{P}_\Omega\{\bm g\}$ denotes the orthogonal projection of $\bm g$ onto $\Omega$, which 
is defined as
\be
\mathcal{P}_\Omega\{\bm g\} = \arg \min_{\widetilde{\bm g} \in \Omega}
\|\widetilde{\bm g}-\bm g\| \: . 
\ee

After tedious but straightforward calculations, it turns out that 
$\nabla_{\widetilde{\bm g}_i} f(\widetilde{\bm G}^{(\kappa)})$ can be expressed as in \eqref{eqn_dbl_y}. 
\begin{figure*}[!t]
\normalsize
\begin{multline}
\label{eqn_dbl_x}
f(\widetilde{\bm G}^{(\kappa)}) = 
\log_2 \left(1  \hspace{-0.5mm} + \frac{\widetilde{\varrho}_i \, \widetilde{\euscr{P}}_i^{(\kappa)} |\widetilde{\bm h}_i^H  \, \widetilde{\bm g}_i^{(\kappa)}|^2}
{\widetilde{\varrho_i} \sum_{\shortstack{\footnotesize $i'=0$ \\ \footnotesize $i' \neq i$}}^{N-1}
\widetilde{\euscr{P}}_{i'}^{(\kappa)} |\widetilde{\bm h}_i^H  \widetilde{\bm g}_{i'}^{(\kappa)}|^2 + \sigma_w^2} \right)
+ \sum_{\shortstack{\footnotesize $i''=0$ \\ \footnotesize $i'' \neq i$}}^{N-1}
 \log_2 \left( 1+\frac{\widetilde{\varrho}_{i''} \, \widetilde{\euscr{P}}_{i''}^{(\kappa)} \,  |\widetilde{\bm h}_{i''}^H  \, \widetilde{\bm g}^{(\kappa)}_{i''}|^2}
 {\widetilde{\varrho}_{i''} \sum_{\shortstack{\footnotesize $i'=0$ \\ \footnotesize $i' \neq i''$}}^{N-1}   \widetilde{\euscr{P}}_{i'}^{(\kappa)} \, 
 |\widetilde{\bm h}_{i''}^H  \, \widetilde{\bm g}^{(\kappa)}_{i'}|^2 + \sigma_w^2} \right) 
\end{multline}
\begin{multline}
\label{eqn_dbl_y}
\nabla_{\widetilde{\bm g}_i} f(\widetilde{\bm G}^{(\kappa)}) = \frac{\widetilde{\varrho}_i \, \widetilde{\euscr{P}}_i^{(\kappa)} \, \widetilde{\bm h}_i \, \widetilde{\bm h}_i^H  \, \widetilde{\bm g}^{(\kappa)}_i}
{\widetilde{\varrho}_i \displaystyle \sum_{i'=0}^{N-1}  \widetilde{\euscr{P}}_{i'}^{(\kappa)} \, 
 |\widetilde{\bm h}_i^H  \, \widetilde{\bm g}^{(\kappa)}_{i'}|^2 + \sigma_w^2}  -
\sum_{\shortstack{\footnotesize $i''=0$ \\ \footnotesize $i'' \neq i$}}^{N-1}
\frac{\widetilde{\varrho}_{i''}^2 \, \widetilde{\euscr{P}}_i^{(\kappa)} \, \widetilde{\euscr{P}}_{i''}^{(\kappa)} \, |\widetilde{\bm h}_{i''}^H  \, \widetilde{\bm g}^{(\kappa)}_{i''}|^2 \, \widetilde{\bm h}_{i''} 
\, \widetilde{\bm h}_{i''}^H  \, \widetilde{\bm g}^{(\kappa)}_i}
{\left[\displaystyle \widetilde{\varrho}_{i''} \sum_{i'=0}^{N-1}  \widetilde{\euscr{P}}_{i}^{(\kappa)} \, 
 |\widetilde{\bm h}_{i''}^H  \, \widetilde{\bm g}^{(\kappa)}_{i'}|^2 + \sigma_w^2\right] \left[\displaystyle 
 \widetilde{\varrho}_{i''} \sum_{\shortstack{\footnotesize $i'=0$ \\ \footnotesize $i' \neq i''$}}^{N-1}
  \widetilde{\euscr{P}}_{i'}^{(\kappa)} \, 
 |\widetilde{\bm h}_{i''}^H  \, \widetilde{\bm g}^{(\kappa)}_{i'}|^2 + \sigma_w^2\right]} 
\end{multline}
\hrulefill
\vspace*{4pt}
\end{figure*}
The step-size $\eta^{(\kappa)}$ is selected using the {\em backtracking} line-search method \cite{Boyd}. 
To prevent the issue of gradient explosion, $\nabla_{\widetilde{\bm g}_i} f(\widetilde{\bm G}^{(\kappa)})$ 
is normalized to unit norm at each iteration.
A remarkable property of the PGA algorithm is that it will
make no progress if $\widetilde{\bm g}^{(\kappa)}_i$ is a stationary point of 
\eqref{eq:prob_G-Gauss} \cite{Boyd}.

\subsubsection{Step $2$: Updating rule of the per-stream power weights}
\label{sec:step-2}

The main challenge in solving \eqref{eq:prob_P-Gauss} 
arises from the 
joint power constraint \eqref{eq:c2} 
imposed on the selected users, rather
than the individual power constraints typically encountered in conventional 
uplink scenarios. Such a difficulty can be circumvented by resorting to the 
downlink-uplink duality \cite{Jindal.2004}, which allows to transform the nonconvex
downlink problem into a convex sum power uplink problem, which is much easier to solve. 
Along this line, we effectively solve problem \eqref{eq:prob_P-Gauss} by implementing the 
{\em sum power iterative waterfilling algorithm} \cite{Goldsmith.2005}, which has been  
also employed in \cite{Di Renzo-ICC} recently. 

Let us employ as initial power policy the average power allocation, i.e., 
$\widetilde{\euscr{P}}_i^{(0)}=\euscr{P}_{\text{tot}}/N$. 
At each iteration, treating the $N$ interference channels  
as their parallel, noninterfering equivalents, the optimal power policy 
is derived by employing the iterative water-filling rule
\be
\widetilde{\euscr{P}}_i^{(\kappa+1)} =\left[\mu - \frac{\widetilde{\varrho_i} \sum_{\shortstack{\footnotesize $i'=0$ \\ \footnotesize $i' \neq i$}}^{N-1} 
\widetilde{\euscr{P}}_{i'}^{(\kappa)} |\widetilde{\bm h}_i^H  \widetilde{\bm g}_{i'}^{(\kappa)}|^2 + \sigma_w^2}{\widetilde{\varrho}_i \, |\widetilde{\bm h}_i^H  \, \widetilde{\bm g}_i^{(\kappa)}|^2} \right]^+
\label{eq:wf-simple}
\ee
for each $i \in \mathcal{N}$, where the water-filling
level $\mu$ is chosen such that 
$\sum_{i=0}^{N-1} \widetilde{\euscr{P}}_{i}^{(\kappa+1)} \leq \euscr{P}_{\text{tot}}$.
It is noteworthy that, to maintain a common water-level, all the $N$
per-stream power weights are simultaneously updated at each iteration, i.e., the $N$
equivalent channels are  simultaneously water-filled.
The iterative steps \eqref{eq:PGA} and \eqref{eq:wf-simple} are repeated 
until a convergence criterion is met, such as when the change in the objective functions 
or the parameters falls below a specified threshold, 
or when a pre-defined iteration limit $\kappa_{\text{max}}$ is reached.

It has been proven in \cite{Goldsmith.2005} that such a simple and highly
intuitive algorithm converges to the sum-rate capacity when $N=2$.
In order to ensure steady convergence to the optimum when $N>2$, it has been proposed 
in \cite{Goldsmith.2005} to introduce a memory in the 
iterative water-filling process. The modified algorithm 
is based on the same basic iterative water-filling rule, but in
each iteration, the updated per-stream powers are now 
a weighted combination of the previous power weights and the new ones generated by the iterative water-filling procedure.
Such a modification has been also implemented in \cite{Di Renzo-ICC}. 
In our optimization, the stability of the powers' update process for 
$N>2$ is ensured by the fact that \eqref{eq:wf-simple} is inserted in a coordinate 
descent method, where the step size of the 
PGA iteration \eqref{eq:PGA} is iteratively shrunk
(i.e., ``backtracked'').
In Section~\ref{sec:numer}, we will numerically validate the convergence
of the  first subproblem.

\subsection{Computation of the
transmission coefficients}
\label{sec:mod-algo}

Once the optimal matrix ${\bm G}^\star$ maximizing
the sum-rate capacity \eqref{eq:prob-1} has been determined, 
after solving the problem in Subsection~IV-\ref{sec:first-prob}
for each subset $\mathcal{K}_N$ and developing the exhaustive search \eqref{eq:scheluding-opt}, 
the next step is to compute 
the optimal transmission coefficients
$\{\pmb{\gamma}^\star_\ell\}_{\ell \in \mathcal{L}}$ of the SIM.
This is done by
solving  equation \eqref{eq:c1} in a {\em least-square} (LS) sense, under the constraints 
\eqref{eq:c3}, \eqref{eq:c4}, and \eqref{eq:c6}. 
Specifically, we minimize the cost function 
\be
\label{eq:c1_mod}
f_2\left(\{\pmb{\gamma}_\ell\}_{\ell \in \mathcal{L}}\right) \eqdef \|{\bm G}- {\bm G}^\star\|^2
\ee
with respect to $\pmb \gamma_1, \pmb \gamma_2, \ldots, \pmb \gamma_L$,
s.t. constraints \eqref{eq:c1}, \eqref{eq:c3}, \eqref{eq:c4}, and \eqref{eq:c6}.

The LS problem \eqref{eq:c1_mod} is solved using the alternating 
 {\em projected gradient descent (PGD)} algorithm \cite{Beck}. This approach alternates optimization with respect to the $L$ metasurface layers.
In particular, the PGD algorithm operates in two main steps. First, it minimizes the cost function 
$f_2\left(\{\pmb{\gamma}_\ell\}_{\ell \in \mathcal{L}}\right)$ with respect to the phases of the PC layers
$\pmb{\phi}_{\ell_1} \eqdef [\phi_{\ell_1,0}, \phi_{\ell_1,1}, \ldots, \phi_{\ell_1,Q-1}]^\trasp$, for $\ell_1 \in \mathcal{L}_{\text pc}$, and the amplitudes of the AC layers
$\pmb{\alpha}_{\ell_2} \eqdef [\alpha_{\ell_2,0}, \alpha_{\ell_2,1}, \ldots, \alpha_{\ell_2,Q-1}]^\trasp$, for $\ell_2 \in \mathcal{L}_{\text ac}$. 
This step ensures compliance with the constraints
\eqref{eq:c1}, \eqref{eq:c3}, and \eqref{eq:c4}.
Second, to fulfil the amplitude constraint \eqref{eq:c6}, the algorithm finds the point within the interval $[\alpha_{\text{min}},\alpha_{\text{max}}]$ that is ``closest'' 
(in the minimum-distance Euclidean sense) to the amplitude determined in the previous step through the gradient descent algorithm.

Starting from the initial points $\pmb{\phi}^{(0)}_{\ell_1}$ and $\pmb{\alpha}^{(0)}_{\ell_2}$, the PGD algorithm iteratively updates the parameters at hand as 
\begin{eqnarray}
\pmb{\phi}^{(\kappa+1)}_{\ell_1}  \hspace{-3mm}&= & \hspace{-2mm} \pmb{\phi}^{(\kappa)}_{\ell_1} - \lambda_{{\phi}_{\ell_1}}^{(\kappa)} \, 
\nabla_{\pmb{\phi}^{(\kappa)}_{\ell_1}}  f_2\left(\{\pmb{\gamma}_\ell\}_{\ell \in \mathcal{L}}\right) 
\label{eq:phi-rule}
\\
\pmb{\alpha}^{(\kappa+1)}_{\ell_2}  \hspace{-3mm}&= &\hspace{-2mm} \mathcal{P}_A \left\{ \pmb{\alpha}^{(\kappa)}_{\ell_2} - 
\lambda_{{\alpha}_{\ell_2}}^{(\kappa)} \, 
\nabla_{\pmb{\alpha}^{(\kappa)}_{\ell_2}} f_2\left(\{\pmb{\gamma}_\ell\}_{\ell \in \mathcal{L}}\right) \right\}
\label{eq:alpha-rule}
\end{eqnarray}
for $\ell_1 \in \mathcal{L}_{\text{pc}}$ and $\ell_2 \in \mathcal{L}_{\text{ac}}$, 
where $\lambda_{{\psi_{\ell_1}}}^{(\kappa)}$ and $\lambda_{{\alpha_{\ell_2}}}^{(\kappa)}$ are the step-sizes for the two update rules
chosen according to the backtracking line-search method \cite{Beck}, 
\be
A \eqdef \left\{\alpha \in \Rset \,: \,  
\alpha_{\text{min}} \le \alpha \le \alpha_{\text{max}} \right\}
\ee
and 
\be
\mathcal{P}_A\{\alpha\} = \arg \min_{\widetilde{\alpha} \in A}
|\widetilde{\alpha}-\alpha| 
\ee
is the orthogonal projection of $\alpha$ onto $A$.
The updates in \eqref{eq:phi-rule} and \eqref{eq:alpha-rule} continue iteratively 
until either the objective function or the parameters fall below a specified threshold, 
or a pre-defined maximum number of iterations $\kappa_{\text{max}}$ is reached.

To compute the gradients in \eqref{eq:phi-rule} and \eqref{eq:alpha-rule}, 
we conveniently express the cost function \eqref{eq:c1_mod} as follows
\barr
f_2\left(\{\pmb{\gamma}_\ell\}_{\ell \in \mathcal{L}}\right) & =
\sum_{i=0}^{N-1}  \|{\bm g}_i-{\bm g}^\star_i\|^2 
\nonumber  \\
& = \sum_{i=0}^{N-1} \left[ {\bm g}_i^\herm \, ({\bm g}_i - {\bm g}^\star_i) -
({\bm g}^{\star}_i)^\herm \, ({\bm g}_i - {\bm g}^\star_i)\right]
\earr
where we have exploited the partitioned structure \eqref{eq:GKN}
and denoted as ${\bm g}^\star_i$ the $i$-th column of the optimal 
beamforming matrix ${\bm G}^\star$, which is obtained from
the iterative algorithm \eqref{eq:PGA} and the solution to \eqref{eq:scheluding-opt}.
Moreover, we observe that, according to \eqref{eq:forward}, the $i$-th column ${\bm g}_i$ of ${\bm G}$ admits the factorization
\be
{\bm g}_i = \bm E_{\ell} \, \diag(\bm b_{\ell,i}) \, \pmb{\gamma}_\ell 
\label{eq:g}
\ee
where the matrix $\bm E_{\ell} \in \Cset^{Q \times Q}$ is {\em extracted} from ${\bm G}$ as $\bm E_{\ell} \eqdef 
\bm \Gamma_L \, \bm W_{L} \, \cdots \bm \Gamma_{\ell+1} \, \bm W_{\ell+1}$ for $\ell \in \mathcal{L}-\{L\}$ and $\bm E_{L} \eqdef \bm I_Q$, whereas $\bm b_{\ell,i}$ is the $i$-th column of 
the matrix $\bm B_\ell \eqdef \bm W_{\ell} \, \bm \Gamma_{\ell-1} \, \bm W_{\ell-1} \cdots \bm \Gamma_{1} \, \bm W_{1} \in \Cset^{Q \times N}$.
After some algebraic manipulations, we obtain
\be
\nabla_{\pmb{\phi}_{\ell_1}}  f_2\left(\{\pmb{\gamma}_\ell\}_{\ell \in \mathcal{L}}\right)  
= 2 \, \Im\left\{ \text{diag}(\pmb{\gamma}^*_{\ell_1}) \left[ \bm A_{\ell_1} \, \pmb{\gamma}_{\ell_1} - \bm v_{\ell_1}  \right] \right\}
\label{eq:grad-max-1}
\ee
for $\ell_1 \in \mathcal{L}_{\text pc}$, and 
\be
\nabla_{\pmb{\alpha}_{\ell_2}}  f_2\left(\{\pmb{\gamma}_\ell\}_{\ell \in \mathcal{L}}\right) 
= 2 \, \Re \left\{ \text{diag}(\pmb{\gamma}^*_{\ell_2}) \left[ \bm A_{\ell_2} \, \pmb{\gamma}_{\ell_2} - \bm v_{\ell_2}  \right] \right\}
\label{eq:grad-max-2}
\ee
for ${\ell_2} \in \mathcal{L}_{\text{ac}}$,  where we have defined
\begin{eqnarray}
\bm A_\ell & \eqdef & \sum_{n=0}^{N-1} \diag(\bm b^*_{\ell,n}) \, \bm E^\herm_{\ell} \, \bm E_{\ell} \, \diag(\bm b_{\ell,n}) 
\label{eq:mat-A}
\nonumber \\
& =& (\bm B^*_\ell \, \bm B^\trasp_\ell) \circ (\bm E^\herm_{\ell} \, \bm E_{\ell} ) \in \mathbb{C}^{Q \times Q} \\
\bm v_\ell & \eqdef & \sum_{n=0}^{N-1} \diag(\bm b^*_{\ell,n}) \, \bm E^\herm_{\ell} \, \widetilde{\bm g}^\star_n 
\nonumber \\
& = &\left[ \bm E^\herm_{\ell} \circ \left( \bm B^*_\ell \, \widetilde{\bm G}^\trasp\right)\right] \bm{1}_Q \in \mathbb{C}^{Q \times 1} \: .
\label{eq:vet-v}
\end{eqnarray}

The study of the convergence of the PGD algorithm requires the introduction of the 
gradient mapping operator, which is an extension of the usual gradient operation. 
We refer to \cite{Beck} for convergence results in terms of the norm of the gradient mapping.

\subsection{Phase discretization for PC layers}
\label{sec:dis-phase}

So far, we have assumed that the phases of the PC layers can assume
any value in the interval $[0, 2 \pi)$. We briefly discuss now
how continuous phases can be transferred into their discrete counterparts
({\em phase discretization process}).

Discrete optimization problems are generally hard to solve efficiently. 
In the context of beamforming, two commonly used methods for solving
discrete optimization problems are integer linear programming (ILP),  for which 
the globally optimal solution  can be obtained by applying the branch-and-bound 
method \cite{Wu.2020}, and  quantization, which involves solving the continuous 
version of the problem and, then, discretizing the obtained solution.
Recently, the use of quantum algorithms has also been proposed for 
beamforming design of reconfigurable intelligent
surfaces \cite{Lim}. The worst-case complexity of ILP 
is exponential over $Q$ due to its fundamental NP-hardness. 
Here,  
we employ quantization to reduce computational complexity, although it does not offer
formal guarantees of optimality.

We implement in Section~\ref{sec:numer} two different quantization strategies.
In the former one, let $\pmb{\phi}^\star_{\ell_1}$ be the
convergence point of the sequence \eqref{eq:phi-rule},
for $\ell_1 \in \mathcal{L}_{\text{pc}}$, such a value 
is quantized to the nearest (in Euclidean distance) value 
$\overline{\pmb{\phi}}^\star_{\ell_1}$ in the
set $\Phi$ defined in Subsection~\ref{sec:model}-A. 
In the latter one, referred to as {\em step-by-step} quantization, 
for $\ell_1 \in \mathcal{L}_{\text{pc}}$, 
the continuous phases obtained through \eqref{eq:phi-rule}
are quantized to the nearest values 
$\overline{\pmb{\phi}}^{(\kappa+1)}_{\ell_1}$ belonging to
$\Phi$ for {\em each} iteration $\kappa$.

\subsection{Computational complexity analysis}
\label{sec:comp-opt}

The design of the SIM implementing optimal beamforming  
consists of four stages: (i) the calculation of the beamforming matrix and the optimal power allocation 
(see Subsection~IV-\ref{sec:first-prob});
(ii) the optimal user group selection in \eqref{eq:scheluding-opt};
(iii) the derivation of the SIM parameters 
(see Subsection~IV-\ref{sec:mod-algo});
and, finally, (iv) the phase discretization process for PC layers.
The last stage  has a negligible implementation cost with 
respect to the other ones and, thus, it is not considered herein.

Determining the optimal user selection in \eqref{eq:scheluding-opt} requires an exhaustive search over the 
entire user set $\mathcal{K}$. The size of the search space is given by the binomial coefficient
$\binom{K}{N}$. Such a brute-force search may be feasible for a relatively small number
of users. For example, with $K=10$ users
and $N=4$ transmit antennas, we have $\binom{10}{4}=210$. 
It should be noted that only the problem
in Subsection~IV-\ref{sec:first-prob} needs to be solved
for each scheduling subset $\mathcal{K}_N \subset \mathcal{K}$. 

For a given user configuration $\mathcal{K}_N$, the computational complexity of the
problem in Subsection~IV-\ref{sec:first-prob} is dominated by  the calculation of the gradient \eqref{eqn_dbl_y}
and by the evaluation of the water-filling rule \eqref{eq:wf-simple}, for each $i \in \mathcal{N}$.  
These operations collectively involve $\mathcal{O}(N^3 \, Q)$
floating point operations (flops) per iteration, when the step size is kept constant.\footnote{Assessing the computational  
burden when using the backtracking line-search method 
for selecting the step size is challenging, since the process 
involves multiple evaluations of the objective function \eqref{eqn_dbl_x}, and the number of these evaluations depends on the specific scenario.}

The computational complexity of the problem discussed in 
Subsection~IV-\ref{sec:mod-algo} is mainly driven by the computation of the 
gradients \eqref{eq:grad-max-1} and \eqref{eq:grad-max-2}, which 
in turn is largely influenced by the calculation of the matrix 
\eqref{eq:mat-A} and the vector \eqref{eq:vet-v}, given a fixed step size.
Such element-wise operations amount to $\mathcal{O}(Q^2)$ flops per iteration.

\begin{figure*}[t]
\centering
\includegraphics[width=1\columnwidth]{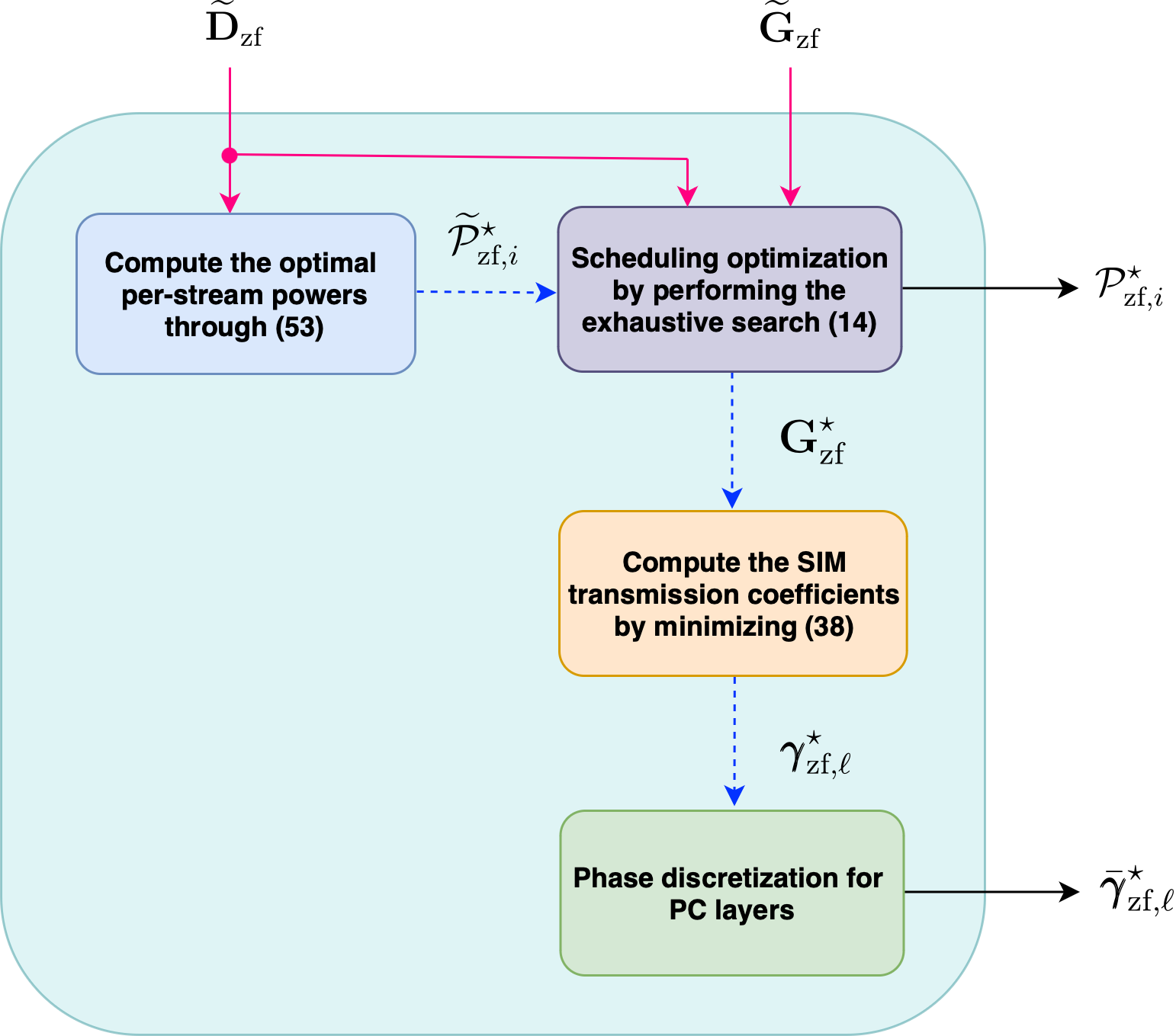}
\caption{Summary of the proposed algorithm maximizing 
the system sum-rate capacity with the ZF constraint.}
\label{fig:fig_3}
\end{figure*}

\section{SIM implementing zero-forcing beamforming}
\label{sec:zfbf}

In this section, we aim to develop a straightforward transmit strategy that is easy to
implement while delivering performance comparable
to the optimal beamforming solutions derived in Section~\ref{sec:opt-bf}.
In particular, we consider a suboptimal beamforming
strategy, referred to as {\em ZF beamforming}, where the
weight vectors are chosen to avoid interference among 
user streams \cite{Goldsmith.2006}. 
Specifically, the matrix $\widetilde{\bm G}$ is designed to solve the 
constrained maximization problem \eqref{eq:prob-1}-\eqref{eq:c5},
with the additional constraints that 
$\widetilde{\bm h}_i^H  \widetilde{\bm g}_{i'}=0$ for $i' \neq i$
and $\widetilde{\bm h}_i^H  \, \widetilde{\bm g}_i=\widetilde{d}_{i}$,
where the real-valued constant $\widetilde{d}_{i}>0$ is introduced
to fulfil the norm constraint \eqref{eq:c2-bis}.
In matrix form, these additional constraints can be concisely written as 
\be
\widetilde{\bm H} \, \widetilde{\bm G} =  \widetilde{\bm D}
\label{eq:zf-constr}
\ee
where  
$\widetilde{\bm D} \eqdef \diag(\widetilde{d}_{0}, \widetilde{d}_{1}, \ldots, \widetilde{d}_{N-1})$
and $\widetilde{\bm H}$ has been defined in \eqref{eq:Htilde}.
ZF beamforming is typically considered power-inefficient since the transmission coefficients of
the SIM are not matched to the user channels. However, 
as it will be shown in Section~\ref{sec:numer}, and in line with the findings in \cite{Goldsmith.2006} and \cite{Ngo.2013}, 
the sum-rate performance of ZF beamforming approaches that of the optimal beamforming solution 
when multiuser diversity \cite{Tse-book} can be exploited or 
in the case of favorable propagation.

Under \eqref{eq:zf-constr}, the cost function \eqref{eq:prob-1} boils down to
\be
\label{eq:prob-2}
\rate_\text{zf} =
\sum_{i =0}^{N-1} \log_2 \left (1  \hspace{-0.5mm} + \frac{\widetilde{\varrho}_i \, \widetilde{\euscr{P}}_i \, \widetilde{d}_{i}^2}
{\sigma_w^2} \right) \: .
\ee
For a given subset $\mathcal{K}_N$ of users, 
the ZF constraint allows to decouple the original suboptimal problem, i.e., maximization of 
the sum-rate capacity \eqref{eq:prob-1} s.t. constraints \eqref{eq:c1}--\eqref{eq:c5} and \eqref{eq:zf-constr},
into two steps:  

\begin{enumerate}[1)]

\item 
Find the solution $\widetilde{\bm G}_{\text{zf}}$ and 
$\widetilde{\bm D}_{\text{zf}}$ of 
the matrix equation \eqref{eq:zf-constr} with respect to $\widetilde{\bm G}$
and the normalization constants
$\widetilde{d}_{0}, \widetilde{d}_{1}, \ldots, \widetilde{d}_{N-1}$,
s.t. constraints \eqref{eq:c1}-\eqref{eq:c6};

\item
Given $\widetilde{\bm D}_{\text{zf}}$,  maximize the cost function \eqref{eq:prob-2}
with respect to the 
per-stream powers $\widetilde{\euscr{P}}_0, \widetilde{\euscr{P}}_1, \ldots, \widetilde{\euscr{P}}_{N-1}$, 
s.t. constraints \eqref{eq:c2} and \eqref{eq:c5}. 

\end{enumerate}

The solution of the two aforementioned steps is derived 
in Subsections~V-\ref{sec:step-1-zf} and V-\ref{sec:step-2-zf}.
Subsequently, the optimal subset of users
is determined by solving \eqref{eq:scheluding-opt}, thus
yielding the beamforming matrix 
$\bm G_\text{zf}^\star$ and the power distribution 
$\{\euscr{P}_{\text{zf}, i}^\star\}_{i \in \mathcal{N}}$.
The last operation  
amounts to finding the optimal transmission coefficients
$\{\pmb{\gamma}^\star_{\text{zf},\ell}\}_{\ell \in \mathcal{L}}$ of the SIM obeying 
constraints \eqref{eq:c1}, \eqref{eq:c3}, \eqref{eq:c4}, and \eqref{eq:c6}. 
This can be carried out by setting up a LS problem similar to \eqref{eq:c1_mod}, 
with ${\bm G}^\star$ replaced by ${\bm G}^\star_{\text{zf}}$.
Such a problem can be solved by resorting to the PGD algorithm, as detailed
in Subsection~IV-\ref{sec:mod-algo}.
Once $\{\pmb{\gamma}^\star_{\text{zf},\ell}\}_{\ell \in \mathcal{L}}$ are obtained, 
the phases of the PC layers need to be discretized according to 
one of the two quantization schemes outlined in 
Subsection~IV-\ref{sec:dis-phase}.

The proposed sum-rate capacity maximization procedure with the ZF constraint is 
summarized in Fig.~\ref{fig:fig_3}. 

\subsection{Computation of the
beamforming matrix}
\label{sec:step-1-zf}

As a first step, we solve the matrix equation \eqref{eq:zf-constr} with respect to
the beamforming matrix $\widetilde{\bm G}$. 
Under the assumption that $\widetilde{\bm H}$ is full-row rank, which necessarily
requires that $N \le Q$, the minimum-norm solution of \eqref{eq:zf-constr} is given by
\barr
\widetilde{\bm G}_{\text{zf}} & = \widetilde{\bm H}^\dag \, \widetilde{\bm D}_{\text{zf}}
\nonumber \\ & =
\widetilde{\bm H}^\herm \, 
(\widetilde{\bm H} \, \widetilde{\bm H}^\herm)^{-1} \, \widetilde{\bm D}_{\text{zf}} \:.
\label{eq:G-zf} 
\earr
The second step consists of determining the diagonal matrix $\widetilde{\bm D}_{\text{zf}}$
that allows fulfillment of the unit-norm constraint \eqref{eq:c2-bis} for each
column of the beamforming matrix $\widetilde{\bm G}_{\text{zf}}$.
After some tedious but straightforward algebraic manipulations, it can be shown that 
\eqref{eq:c2-bis} is fulfilled by choosing 
\be
\widetilde{d}_{\text{zf},i} =  
\frac{1}{[(\widetilde{\bm H} \, \widetilde{\bm H}^\herm)^{-1/2}]_{i,i}} \:,
\quad \text{for $i \in \mathcal{N}$} \:.
\label{eq:dtilde}
\ee

\begin{figure*}[t]
\centering
\includegraphics[width=0.8\linewidth]{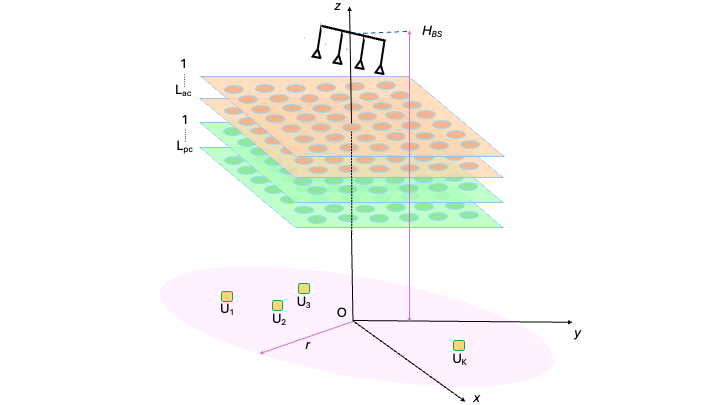} 
\caption{Simulation setup of the downlink multiuser system.}
\label{fig:fig_4}
\end{figure*}

\subsection{Computation of the per-stream power policy}
\label{sec:step-2-zf}

Given the diagonal entries of $\widetilde{\bm D}_{\text{zf}}$ as in \eqref{eq:dtilde}, 
the optimal power allocation $\{\widetilde{\euscr{P}}^\star_{\text{zf},i}\}_{i \in \mathcal{N}}$
can be explicitly found \cite{Tse-book} as 
\be
\widetilde{\euscr{P}}^\star_{\text{zf},i} = \left(\frac{1}{\mu_\text{zf}} - \frac{\sigma_w^2}{\widetilde{\varrho}_i \, \widetilde{d}_{\text{zf},i}^2} \right)^+
\label{eq:wf-zf}
\ee
for $i \in \mathcal{N}$, with the constant $\mu_\text{zf}$ chosen such that the power constraint \eqref{eq:c2} is met
\be
\sum_{i=0}^{N-1} \left(\frac{1}{\mu_\text{zf}} - \frac{\sigma_w^2}{\widetilde{\varrho}_i \, \widetilde{d}_{\text{zf},i}^2} \right)^+
= \euscr{P}_{\text{tot}} \: .
\label{eq:wf-zf-constr}
\ee
The optimal power policy in the case of ZF beamforming is therefore the standard 
waterfilling or waterpouring strategy.

\subsection{Computational complexity analysis}
\label{sec:comp-zf}

The design of the SIM implementing ZF beamforming  
consists of four stages: 
(i) calculation of the beamforming matrix and the optimal power allocation 
(see Subsections~V-\ref{sec:step-1-zf} and V-\ref{sec:step-2-zf});
(ii) optimal user group selection in \eqref{eq:scheluding-opt};
(iii) derivation of the SIM parameters;
and, finally, (iv) phase discretization process for PC layers.
Compared to the optimal case, 
the only significant implementation difference is represented by the first stage (i),
which is not iterative in the ZF case.

The calculation of \eqref{eq:G-zf}-\eqref{eq:dtilde} relies on the 
inversion of the matrix $\widetilde{\bm H} \, \widetilde{\bm H}^\herm$, 
which entails $\mathcal{O}(N^3)$ flops if batch algorithms are used,\footnote{Iterative 
algorithms can be exploited in order to directly
evaluate $(\widetilde{\bm H} \, \widetilde{\bm H}^\herm)^{-1}$,
providing $\mathcal{O}(N^2)$ flops for each iteration.}
while the water-filling algorithm \eqref{eq:wf-zf}-\eqref{eq:wf-zf-constr}
has a very low implementation cost regardless of $Q$. 
It is worth noticing that the complexity of the corresponding step (i)
in the case of optimal beamforming 
(see Subsection~IV-\ref{sec:comp-opt}) is much higher and requires
multiple iterations.

\begin{figure}[t]
\centering
\includegraphics[width=\linewidth]{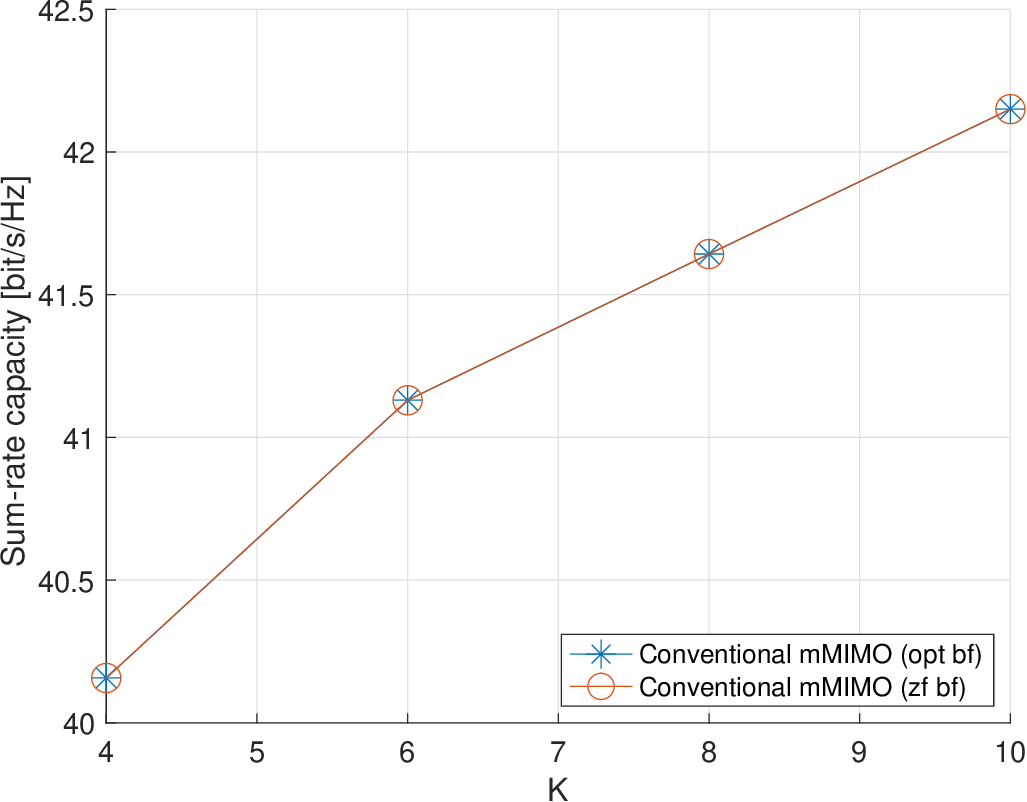} 
\caption{Sum-rate capacity versus number of users ($Q=49$).}
\label{fig:fig_5}
\end{figure}

\section{Monte Carlo numerical results}
\label{sec:numer}

In this section, we present Monte Carlo simulations to validate the 
proposed SIM designs and assess the sum-rate capacity of 
the considered SIM-aided multiuser downlink.

\begin{figure*}[!t]
\begin{minipage}[b]{8.8cm}
\centering
\includegraphics[width=\linewidth]{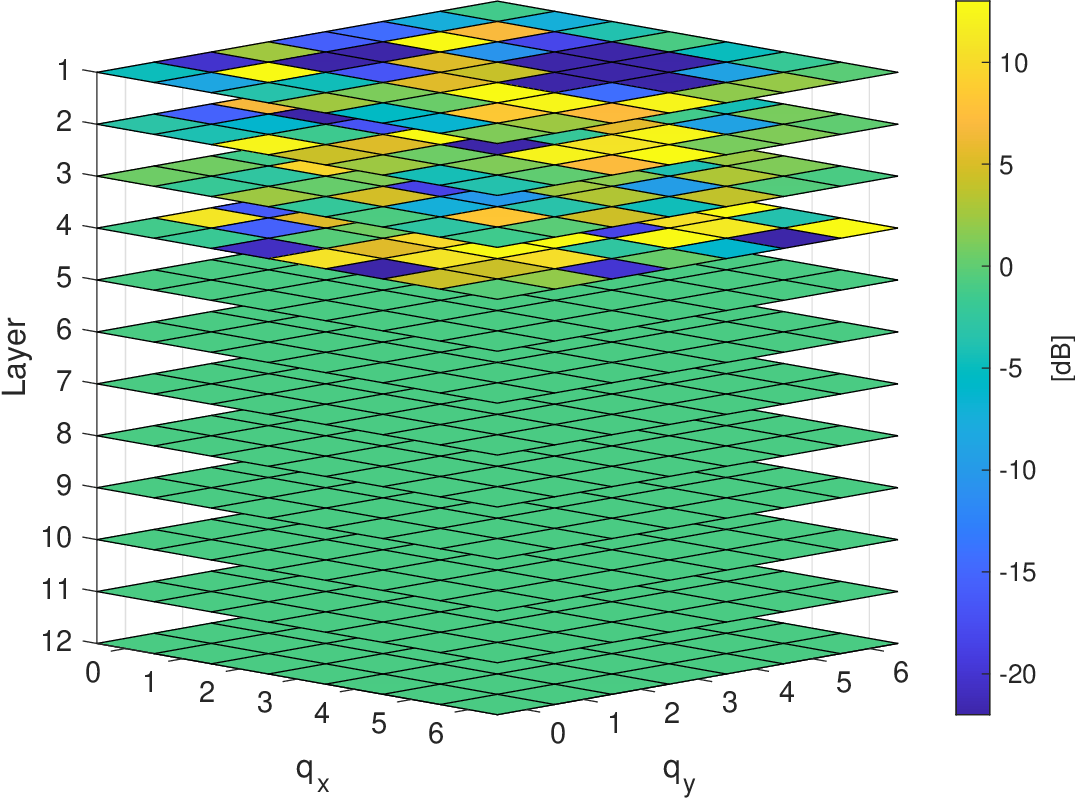}
\end{minipage}
\begin{minipage}[b]{8.8cm}
\centering
\includegraphics[width=\linewidth]{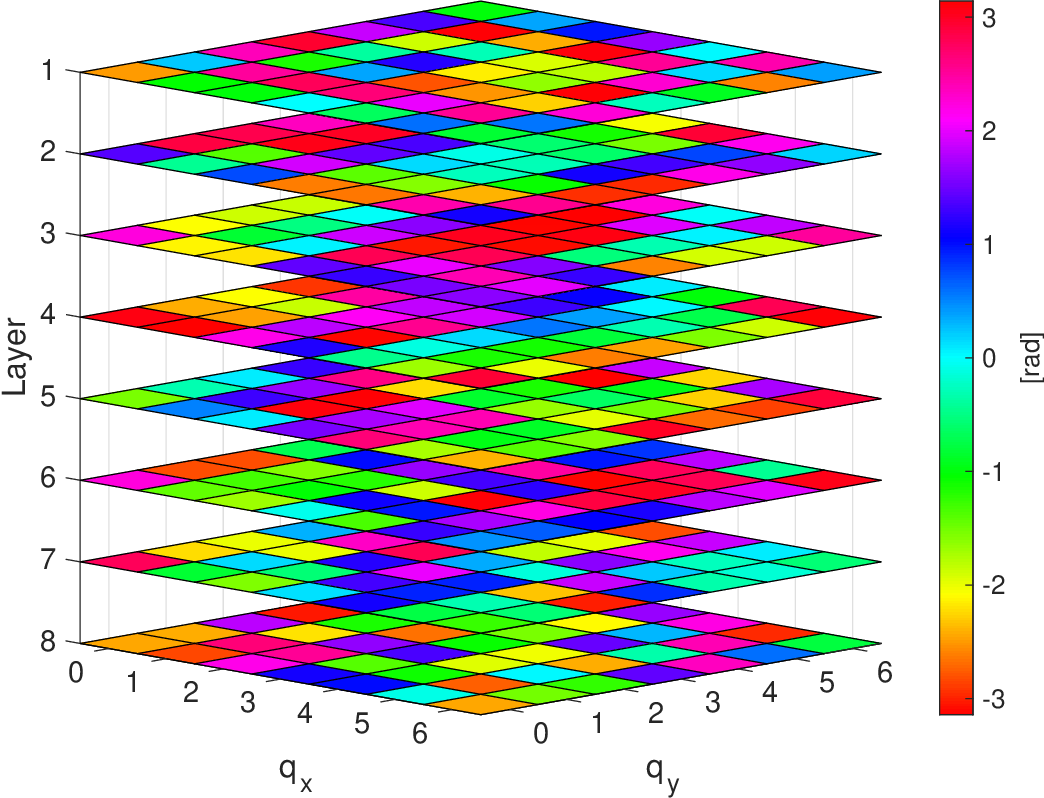}
\end{minipage}
\caption{Heatmap of amplitude response of PC-and-AC SIM (left, $L_{\text{pc}}=8$, $L_{\text{ac}}=4$, $Q=49$, and $K=8$) 
and phase response of PC-only SIM (right, $L_{\text{pc}}=8$, $L_{\text{ac}}=0$, $Q=49$, and $K=8$).}
\label{fig:fig_6}
\end{figure*}
\begin{figure*}[!t]
\begin{minipage}[b]{8.8cm}
\centering
\includegraphics[width=\linewidth]{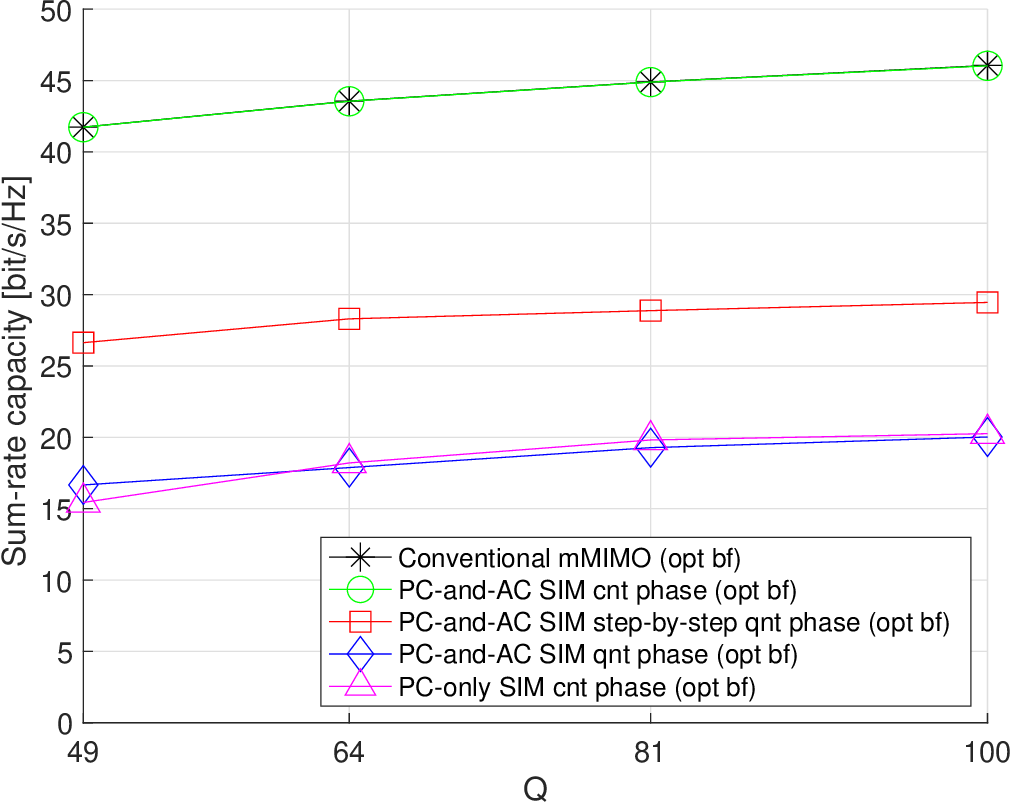}
\end{minipage}
\begin{minipage}[b]{8.8cm}
\centering
\includegraphics[width=\linewidth]{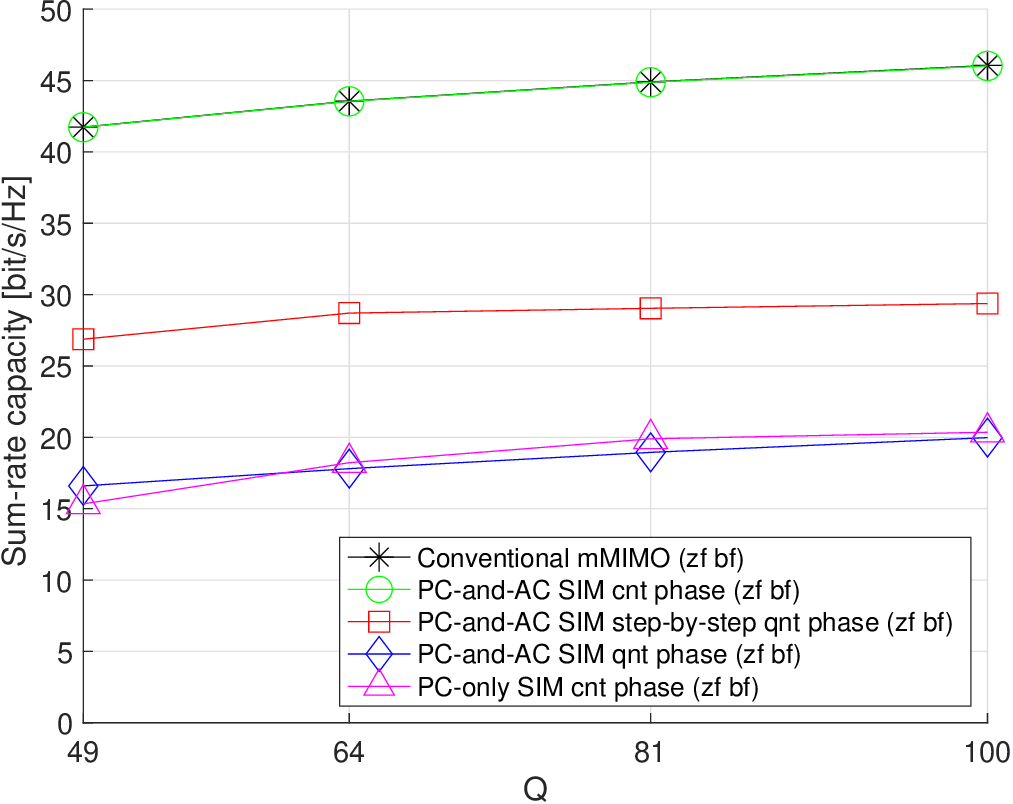}
\end{minipage}
\caption{Sum-rate capacity versus number of meta-atoms for optimal (left) and ZF (right) beamforming
($L_{\text{pc}}=8$ and $L_{\text{ac}}=4$, the signal from the RF chain first passes through the
AC layers and subsequently through the PC layers, $K=8$, and $b=3$).
}
\label{fig:fig_7}
\end{figure*}
\begin{figure*}[!t]
\begin{minipage}[b]{8.8cm}
\centering
\includegraphics[width=\linewidth]{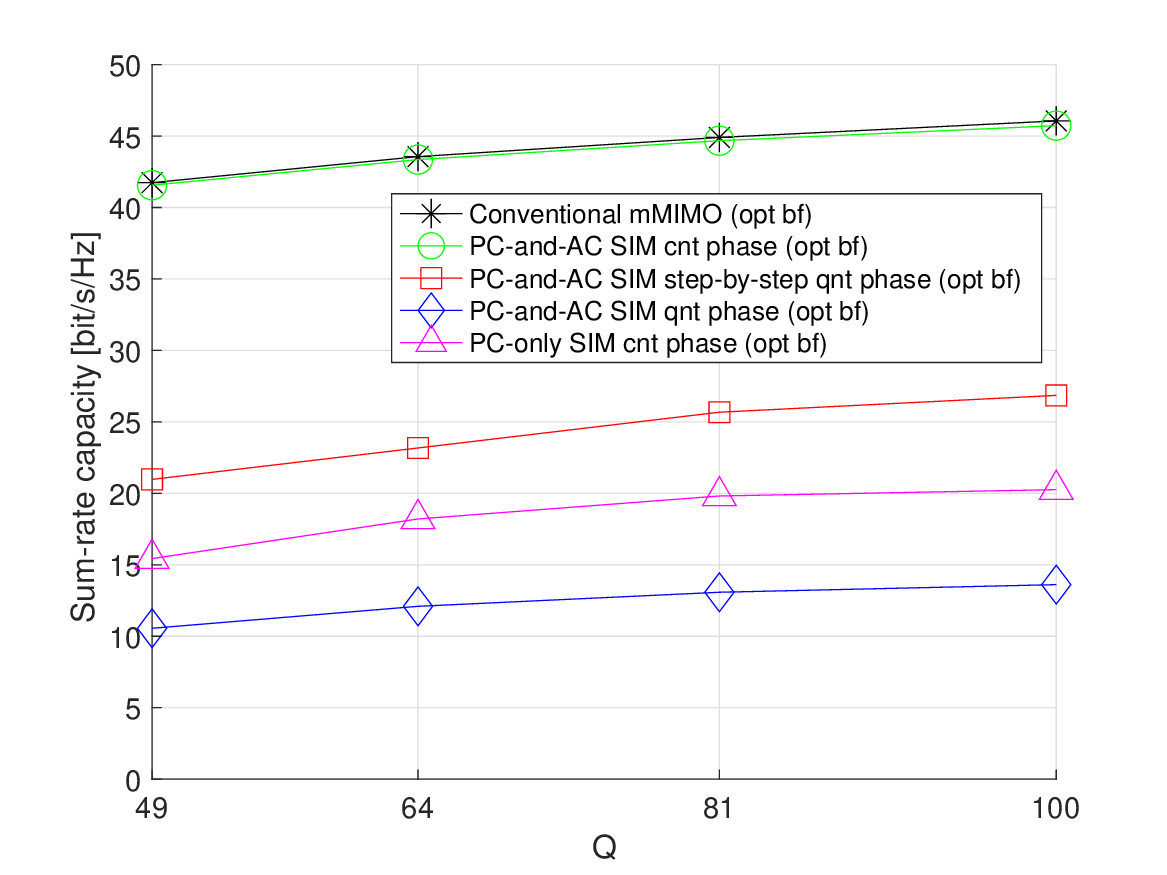}
\end{minipage}
\begin{minipage}[b]{8.8cm}
\centering
\includegraphics[width=\linewidth]{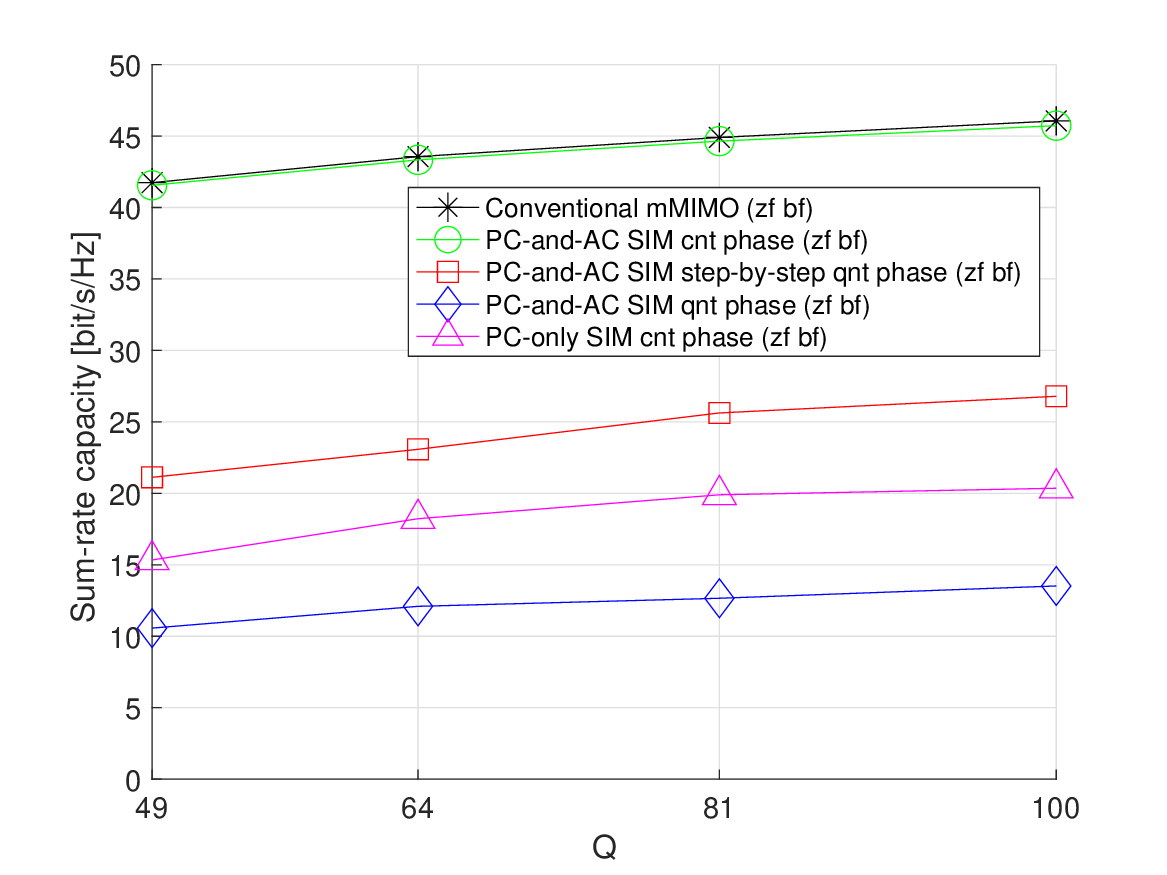}
\end{minipage}
\caption{Sum-rate capacity versus number of meta-atoms for optimal (left) and ZF (right) beamforming
($L_{\text{pc}}=8$, $L_{\text{ac}}=4$, PC and AC layers are interlaced, $K=8$, and $b=3$).
}
\label{fig:fig_8}
\end{figure*}
\begin{figure*}[!t]
\begin{minipage}[b]{8.8cm}
\centering
\includegraphics[width=\linewidth]{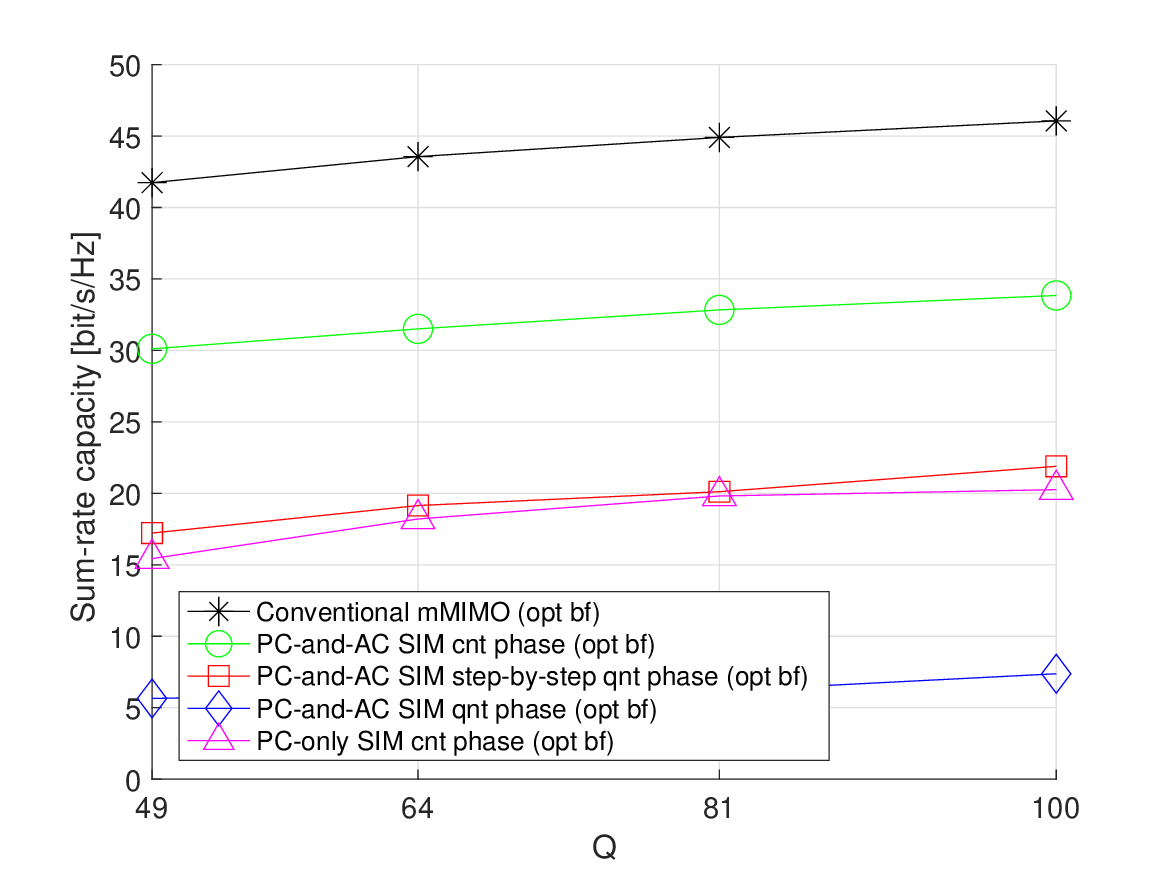}
\end{minipage}
\begin{minipage}[b]{8.8cm}
\centering
\includegraphics[width=\linewidth]{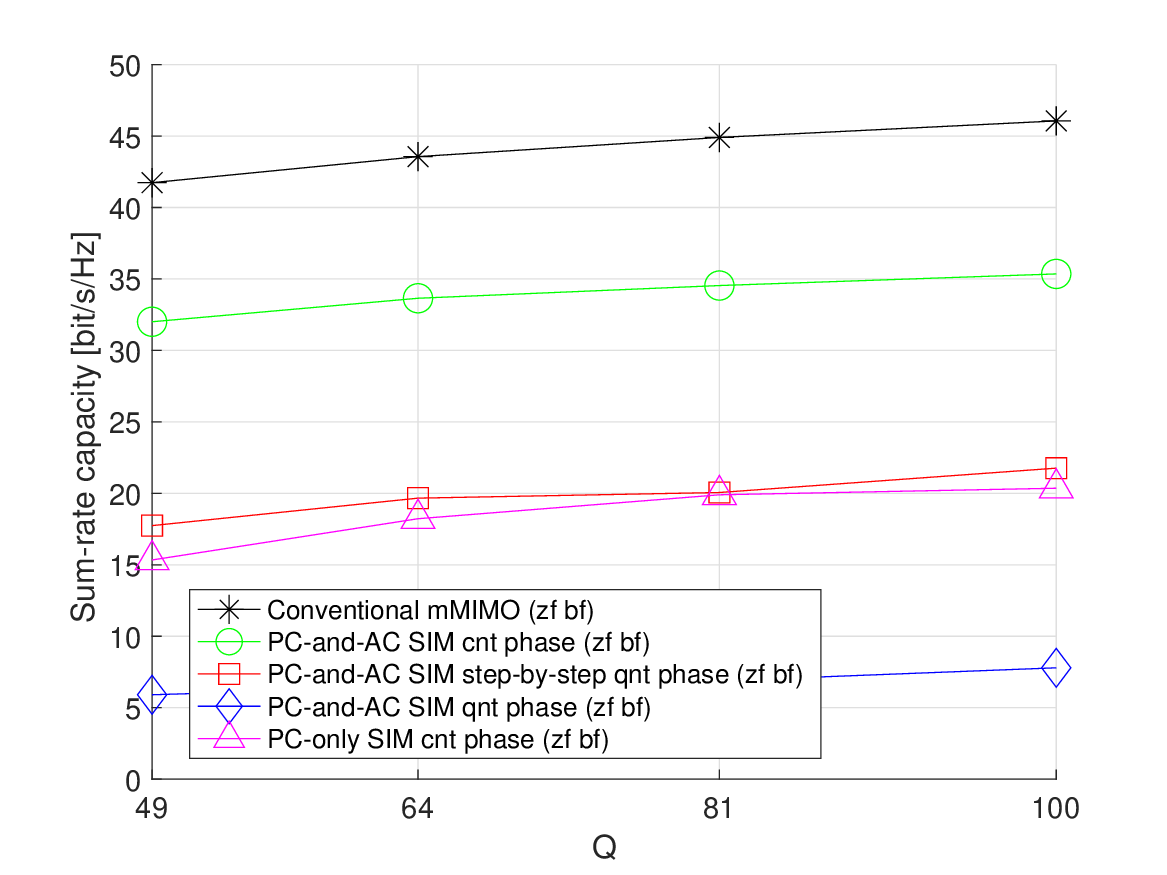}
\end{minipage}
\caption{Sum-rate capacity versus number of meta-atoms for optimal (left) and ZF (right) beamforming
($L_{\text{pc}}=8$, $L_{\text{ac}}=4$, the signal from the RF chain first passes through the
PC layers and subsequently through the AC layers, $K=8$, and $b=3$).
}
\label{fig:fig_9}
\end{figure*}
\begin{figure*}[!t]
\begin{minipage}[b]{8.8cm}
\centering
\includegraphics[width=\linewidth]{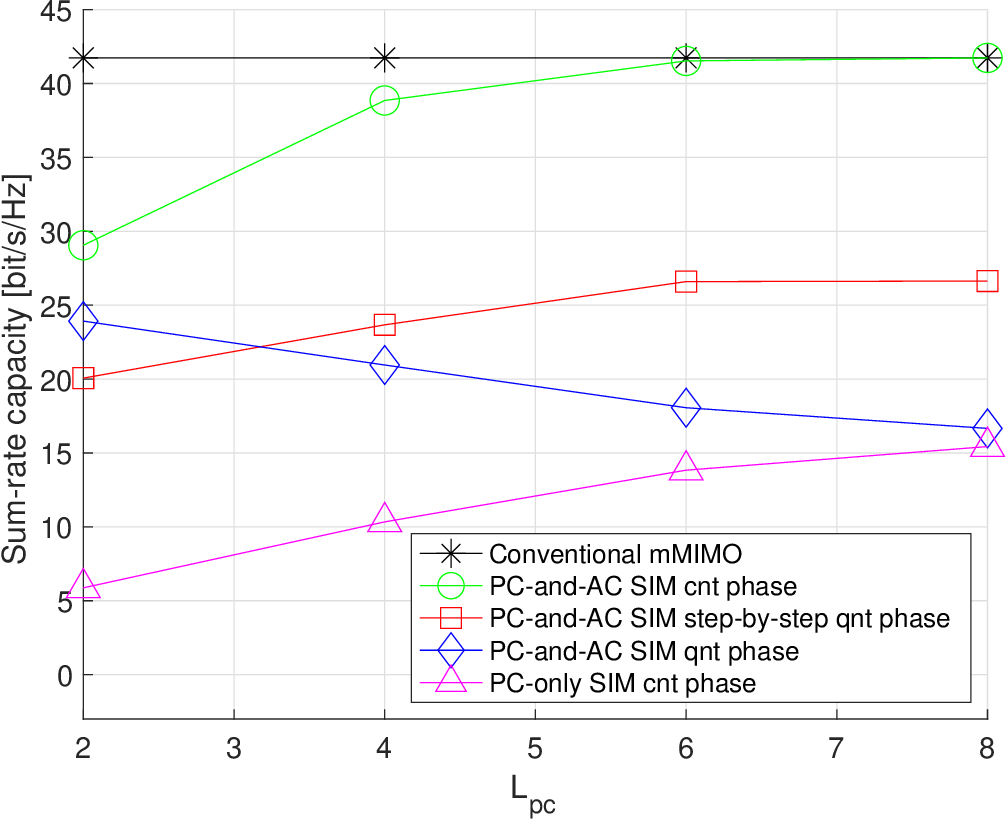}
\end{minipage}
\begin{minipage}[b]{8.8cm}
\centering
\includegraphics[width=\linewidth]{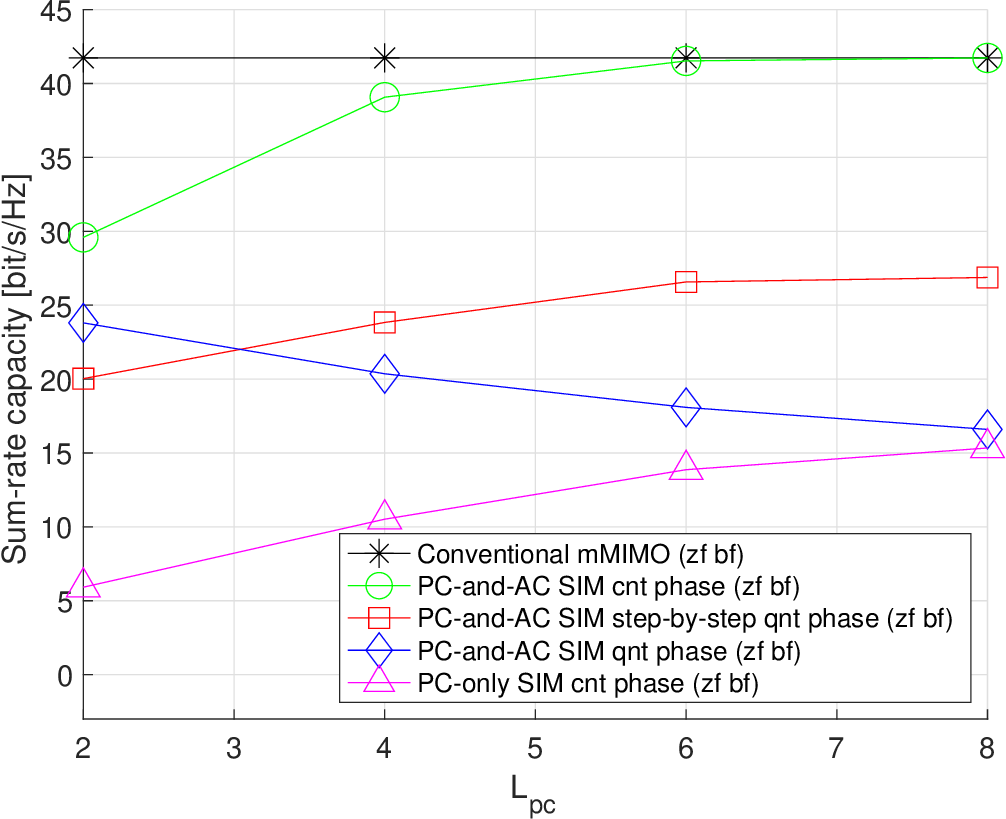}
\end{minipage}
\caption{Sum-rate capacity versus number of PC layers for optimal (left) and ZF (right) beamforming
($L_{\text{ac}}=4$, $Q=49$, $K=8$, and $b=3$).
}
\label{fig:fig_10}
\end{figure*}
\begin{figure*}[!t]
\begin{minipage}[b]{8.8cm}
\centering
\includegraphics[width=\linewidth]{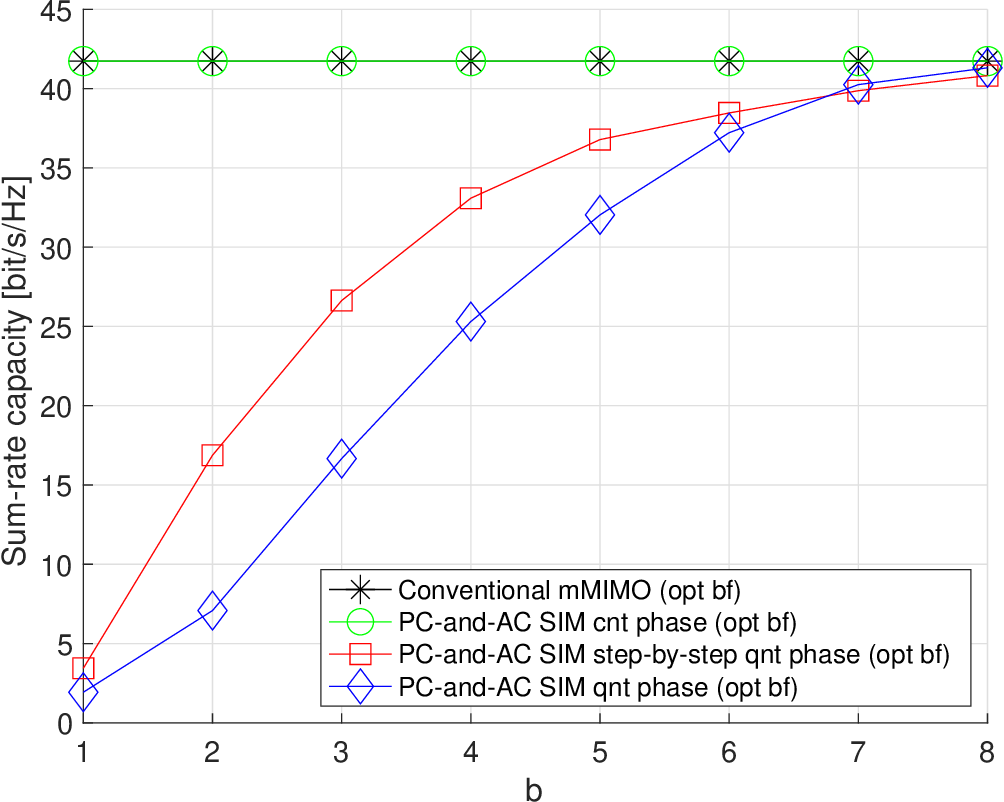}
\end{minipage}
\begin{minipage}[b]{8.8cm}
\centering
\includegraphics[width=\linewidth]{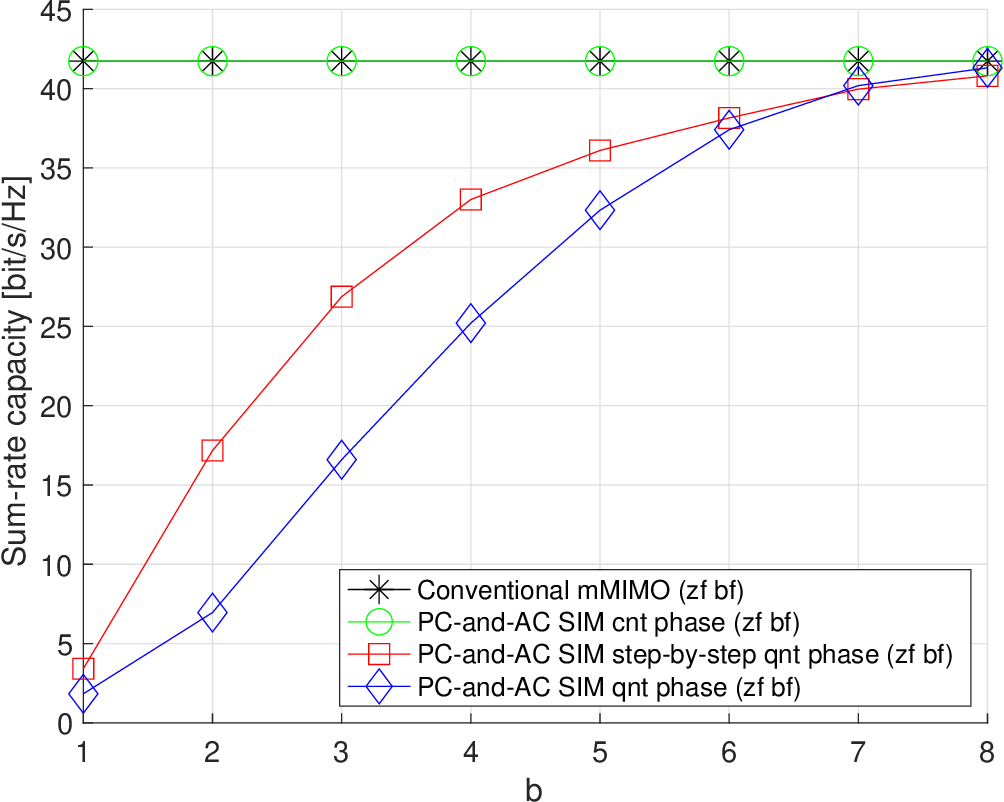}
\end{minipage}
\caption{Sum-rate capacity versus number of bits for optimal (left) and ZF (right) beamforming
($L_{\text{pc}}=8$, $L_{\text{ac}}=4$, $Q=49$, and $K=8$).
}
\label{fig:fig_11}
\end{figure*}
\begin{figure*}[!t]
\begin{minipage}[b]{8.8cm}
\centering
\includegraphics[width=\linewidth]{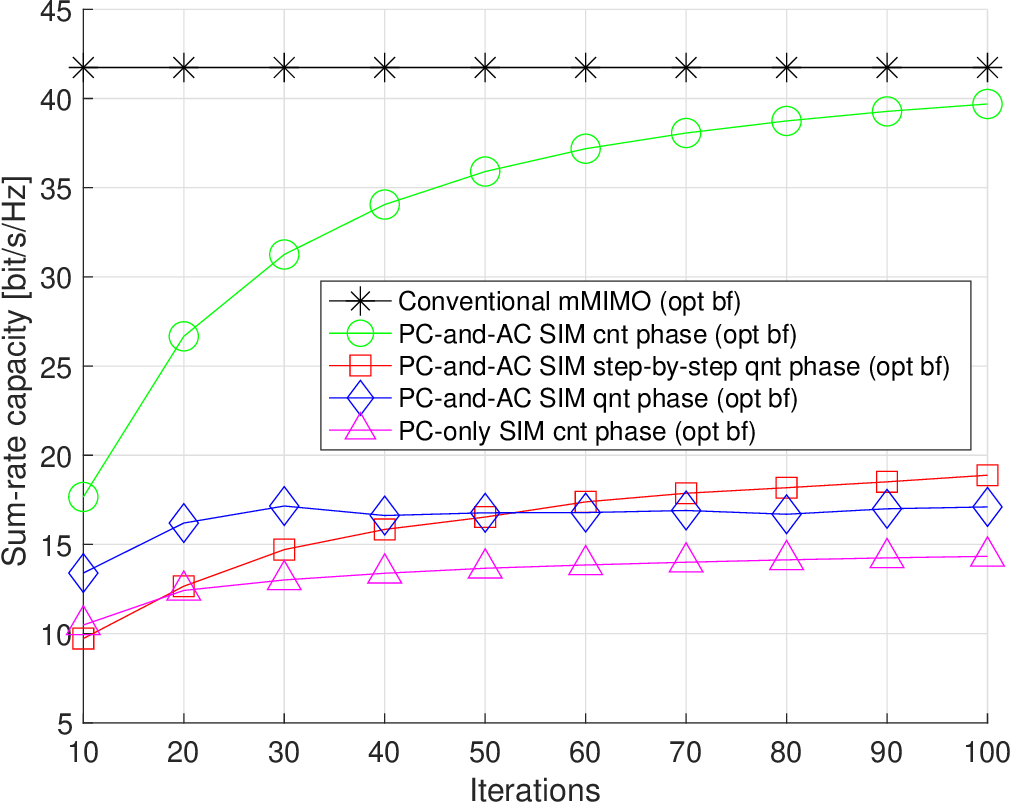}
\end{minipage}
\begin{minipage}[b]{8.8cm}
\centering
\includegraphics[width=\linewidth]{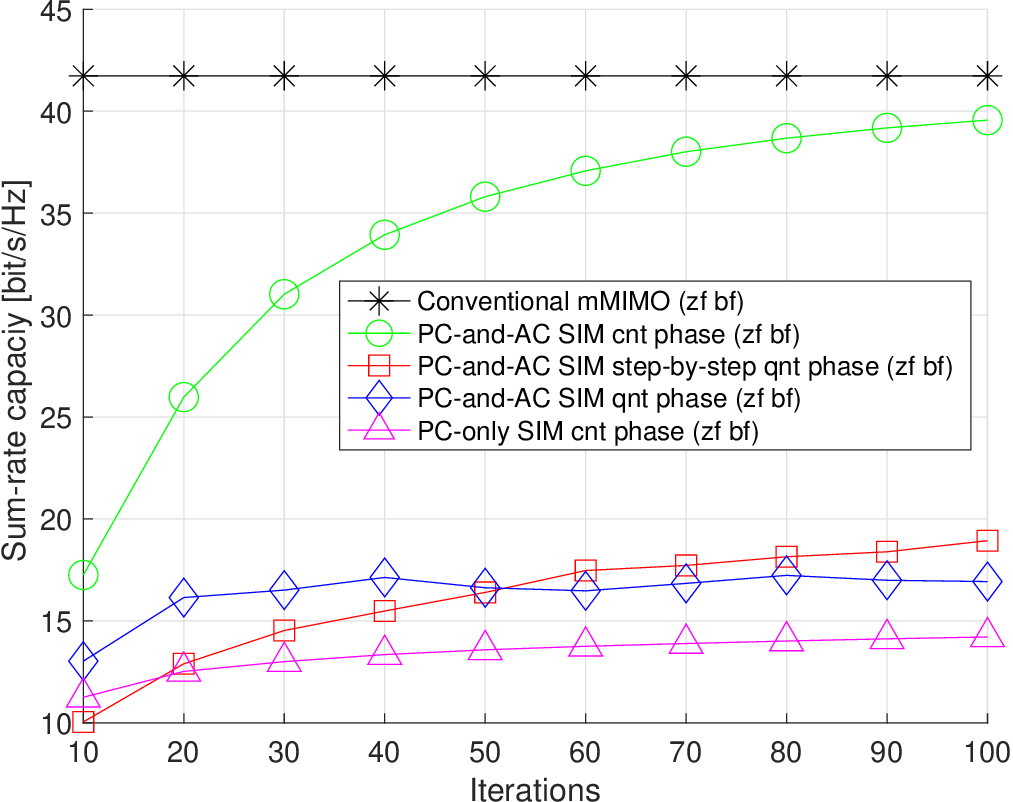}
\end{minipage}
\caption{Sum-rate capacity versus number of iterations for optimal (left) and ZF (right) beamforming
($L_{\text{pc}}=8$, $L_{\text{ac}}=4$, $Q=49$, and $K=8$).
}
\label{fig:fig_12}
\end{figure*}

As illustrated in Fig.~\ref{fig:fig_4}, we consider a three-dimensional Cartesian system, wherein the BS 
is located at $(0,0,H_{\text{BS}})$, with $H_{\text{BS}} = 10$ m, whereas the positions of the $K$ users 
are generated as random variables uniformly distributed within a circular area lying in the $xy$-plane and 
centered in $(0,0,0)$, whose radius is $r=10$ m. The number of users is set to $K=8$, unless 
otherwise specified.
The system operates at a carrier frequency $f_0=28$ GHz, with a transmission bandwidth of $10$ MHz and noise power spectral
density equal to $-174$ dBm/Hz for all the users. The available power budget $\euscr{P}_{\text{tot}}$ at the BS, including the transmit array gain, is fixed to 
$15$ dBm. 
The BS is equipped with a uniform linear array aligned along the $x$-axis, 
consisting of $N=4$ antennas with half-wavelength spacing. The SIM is composed of $L=L_{\text{pc}}+L_{\text{ac}}$ 
layers spaced $s=5 \lambda/L$ apart. Unless otherwise specified, 
we set $L_{\text{pc}}=8$ and $L_{\text{ac}}=4$
for the PC-and-AC SIM configuration, by considering the case 
in which the signal from the RF chain first passes through 
the AC layers and subsequently through the PC layers, 
whereas  
$L_{\text{pc}}=8$ and $L_{\text{ac}}=0$
in the case of PC-only SIM.
We consider three different versions of PC-and-AC SIM
(see Subsection~IV-\ref{sec:dis-phase}): 
in the first version, the phases of the PC layers can assume
any value in the interval $[0, 2 \pi)$ (referred to as ``cnt phase");
in the second version, the phases of the PC layers
are quantized after the convergence of the sequence \eqref{eq:phi-rule}
(referred to as ``qnt phase"); 
in the third version, the phases of the PC layer are quantized step-by-step
(referred to as ``step-by-step qnt phase").
For PC-only SIM, we report the best possible performance when the 
phases of the transmission coefficients are not quantized.
The spacing between the BS array and the first layer of the SIM is $\sigma=s$.
Unless otherwise specified, each layer 
comprises $Q=Q_x \times Q_y=49$ meta-atoms, with $Q_x=Q_y=7$, whose inter-element  
spacing and size are $d_{\text{RIS}} = \lambda/2$ and $A_{\text{meta}} = \lambda^2/4$, respectively.
Moreover, the number of coding bits for the phase values of PC layers is $b=3$, leading to 
$M=8$ possible phase values for PC layers.
The PC layers have a constant 
transmittance $\alpha_\text{pc}=0.9$. 
In line with the $\text{D}^2$NN platform implemented in \cite{Liu.2022}, 
the amplitude responses of AC layers obey 
$\alpha_{\ell,q} \in [\alpha_{\text{min}}, \alpha_{\text{max}}]$,
with $\alpha_{\text{min}}=-22$ dB and 
$\alpha_{\text{max}}=13$ dB, for $\ell \in \mathcal{L}_{\text{ac}}$ and $q \in \mathcal{Q}$.
The receive antenna gain is set to $0$ dBi for each user equipment.

The entries of the user channel $\bm h_k$ are generated as i.i.d. circularly-symmetric complex Gaussian random variables, with zero
mean and unit variance, for each $k \in \mathcal{K}$. 
Regarding the path-loss model adopted in Subsection~II-\ref{sec:model-user}, the reference distance is 
$d_0 = 1$ m and the path-loss exponent is set equal to $\eta=3.5$.
The maximum number of iterations for the iterative algorithms is $\kappa_\text{max}=1000$.
All the results are obtained by averaging over $200$ independent realizations of 
channels, user positions, and noise samples.

As baseline transmission technology, we consider 
the conventional massive MIMO (mMIMO) scheme, where the $N$ multiple
data streams are first linearly precoded 
and then fed to the corresponding $Q \gg N$ transmit antennas. 
We recall that
SIM-based schemes exhibit key advantages in comparison to their mMIMO 
conventional counterparts \cite{Hanzo}, such as improved computational efficiency
(i.e., ultrafast computational speed, parallel computational capability,
and reduced computational complexity), simplified hardware architecture
(e.g., low-resolution DAC/ADC and reduced number of RF chains), and 
reduced energy consumption. 
Fig.~\ref{fig:fig_5} illustrates 
the performance of a mMIMO scheme 
as a function of the number of
users,  where
the $4 \times 49$ precoding matrix is optimized either to 
achieve maximum sum-rate capacity according to
\eqref{eq:PGA} and \eqref{eq:wf-simple} (referred to as 
``opt bf") or to ensure no interference among user
streams through \eqref{eq:G-zf} and 
\eqref{eq:dtilde} (referred to as  ``zf bf"). 
It is observed that the sum-rate capacity achieved by ZF beamforming 
is basically equal to that of the optimal beamforming. 
Such a result comes from the fact that, in the setting at hand, 
the number of antennas is greater than the number of users.
In this operative scenario, favorable propagation occurs \cite{Ngo.2013}, which makes 
the channel directions of the users approximately orthogonal, 
in which case ZF beamforming is nearly optimal.
On the other hand, when the number $K$ of users is greater than $Q$, ZF beamforming 
is suboptimal. However, in this latter case, 
the sum-rate capacity of the ZF beamforming 
approaches the performance of the optimal beamforming
under large $K$ \cite{Goldsmith.2006}, due to the 
multiuser diversity effect \cite{Tse-book}.

Before analyzing the sum-rate performance of the proposed SIM-based 
structures in detail, Fig.~\ref{fig:fig_6} shows the heatmaps of 
the transmission responses of two different SIM configurations
implementing optimal beamforming.
The heatmap on the left-side hand of Fig.~\ref{fig:fig_6} displays the magnitude of 
the individual transmission coefficients of  
PC-and-AC SIM with $8$ PC layers and 
$4$ AC layers. It is interesting to observe that the optimized values
of the amplitude coefficients of the AC layers span the 
whole range of values from 
$\alpha_{\text{min}}=-22$ dB to  
$\alpha_{\text{max}}=13$ dB, thus providing a dynamic modulation range of $35$ dB.
On the other hand, the heatmap on the right-side hand of Fig.~\ref{fig:fig_6} 
shows the phase of the individual transmission coefficients of  
PC-only SIM with $8$ PC layers, which assumes 
any value in the interval $[-\pi, \pi)$, i.e., without phase discretization. 
This heatmap shows that the optimized phases of PC-only SIM exhibit 
rapid changes within a single layer, as well as from one layer
to another.  

\subsection{Performance as function of the number of meta-atoms for 
different arrangements of the SIM layers}

In Figs.~\ref{fig:fig_7}, \ref{fig:fig_8}, and \ref{fig:fig_9}, we 
show the sum-rate capacity of the different transmit schemes under
comparison as a function of the number $Q$ of meta-atoms in each layer, for three
different arrangements of the SIM layers: (i) in Fig.~\ref{fig:fig_7}, 
the signal from the RF chain first passes through the
AC layers and subsequently through the PC layers (RF-AC-PC arrangement); 
(ii) in Fig.~\ref{fig:fig_8}, the PC and AC layers are interlaced;
(iii) in Fig.~\ref{fig:fig_9},  the signal from the RF chain first passes through the
PC layers and subsequently through the AC layers (RF-PC-AC arrangement).
It is evident that the RF-AC-PC arrangement is the best configuration
in terms of sum-rate capacity, for the considered number of iterations. In particular, 
the schemes using PC-and-AC SIM exhibit a noticeable performance degradation
in the case of the RF-PC-AC arrangement. The reason for such different
behaviors of the three configurations under comparison is basically due to the fact that
the positions of the AC/PC layers affects the convergence rate
of the PGD algorithm (i.e., the number of iterations required for the algorithm
to converge to its steady state value), which is used for the SIM optimization 
(see Subsection~IV-\ref{sec:mod-algo}). The highest convergence rate is achieved
when the AC layers are positioned before the PC ones.

The performance of all schemes slowly improves for increasing values of $Q$, by showing 
a marked saturation effect for SIM-based solutions. 
Moreover, SIM implementing ZF beamforming exhibit almost the same performance 
of the SIM mimicking optimal beamforming for all the considered values of $Q$,
but with a reduced computational complexity. 
Remarkably, PC-and-AC SIM ``cnt phase" perform comparably to 
the corresponding mMIMO scheme, while requiring a substantial 
lower cost and overall energy consumption, and, at the same time, they significantly 
outperform PC-only SIM ``cnt phase". Quantitatively speaking,
compared to PC-only SIM ``cnt phase", the rate of PC-and-AC SIM ``cnt phase"
in Fig.~\ref{fig:fig_7} is improved of approximately $25$ bit/s/Hz. 
The phase discretization process in PC layers negatively affects
the performance of PC-and-AC SIM, highlighting that 
step-by-step quantization is a more effective approach. 
Specifically, when $Q=64$, PC-and-AC SIM ``step-by-step qnt phase" 
achieves in Fig.~\ref{fig:fig_7} a rate increase of approximately 
$10$ bit/s/Hz compared to the 
PC-only SIM ``cnt phase".

\subsection{Performance as function of the number of PC layers}
Fig.~\ref{fig:fig_10} compares the sum-rate capacity of the different transmit schemes of
interest as a function of the number $L_\text{pc}$ of PC layers. 
The performance of all SIM-based schemes  improves rapidly with increasing values of $L_\text{pc}$, 
except for the PC-and-AC SIM ``qnt phase", which exhibits deteriorating performance as $L_\text{pc}$ increases.
Such a decline in performance is attributed to
the {\em error propagation phenomenon} in
multi-layer structures. In this scenario, quantization errors in one layer adversely affect the accuracy of quantization in subsequent layers.
This experiment confirms the clear superiority in rate performance of 
the PC-and-AC SIM ``cnt phase" and PC-and-AC SIM ``step-by-step qnt phase" compared to the
 PC-only SIM ``cnt phase". 

\subsection{Performance as function of the number of bits for the phase
values of PC layers}

We investigate in Fig.~\ref{fig:fig_11} the performance of the proposed
PC-and-AC SIM as a function of  the number $b$ of coding bits 
for the phase values of PC layers. 
In this figure, the performance of 
PC-and-AC SIM ``cnt phase" is unaffected by $b$ since 
the phases of the PC layers for this scheme can assume any
value in the interval $[0, 2 \pi)$ (i.e., the phases of 
the PC layers are not quantized).
As anticipated, the performance of structures utilizing  
phase discretization improves significantly as
the number of quantization levels increases. Remarkably,
all three  
versions of the PC-and-AC SIM achieve similar sum-rate capacities when $b=8$. This  experiment 
highlights that  phase quantization  is
considerably more critical for SIM compared to single metasurfaces, due 
to the aforementioned error propagation phenomenon.

\subsection{Performance analysis as function of the number of iterations}

Fig.~\ref{fig:fig_12} illustrates the sum-rate capacity of the various transmit schemes  as a function of the number of iterations of the iterative 
algorithms used for designing the SIM. 
The performance of the PC-and-AC SIM ``cnt phase" steadily 
improves with the number of iterations, demonstrating an
increase of approximately $20$ bit/s/Hz for an order-of-magnitude 
increase in iterations.  
In contrast,  the rate of PC-only SIM ``cnt phase" 
increases more modestly, ranging from approximately $10$ to $15$ bit/s/Hz. Additionally, it is evident
that the phase quantization process substantially impedes
the convergence rate of both 
PC-and-AC SIM ``step-by-step qnt phase" and 
PC-and-AC SIM ``qnt phase".

\section{Conclusions and directions for future work}
\label{sec:concl}

We have developed an optimization framework to design 
SIM composed of both PC and AC layers for enhancing 
the performance of a downlink multiuser system. 
Both optimal beamforming, which maximizes the sum-rate capacity,   
and suboptimal ZF beamforming, which enforces no interference 
among user streams, have been considered. 
We have showed that judiciously reconfiguring both amplitude and phase responses
of the SIM allows to achieve a significant performance 
gain with respect to conventional PC-only SIM. 
In particular, when the number of meta-atoms per layer is sufficiently
greater than the number of users, SIM mimicking ZF beamforming achieve a
sum rate close to the optimal rate promised by SIM
implementing optimal beamforming, but with a much lower complexity. 

In this study, we have focused on SIM composed of layers
where only spatial modulation is employed during each channel coherence time. An important extension would be to explore
the introduction of temporal  modulation for the transmission 
responses of the layers.
Additionally, further investigation is needed into the potential benefits of incorporating stable
nonlinear amplifiers within the meta-atoms.

\appendix
\label{sec:appendix}

This appendix aims to briefly review the mathematical model for the propagation
of EM waves between diffractive layers, which is essential for 
modeling and optimizing SIM.

According to Huygens' geometrical construction, every point on a wavefront acts as a secondary source emitting spherical wavelets. The wavefront at any subsequent instant can be seen as the envelope of
these wavelets. 
Fresnel extended this construction by incorporating Young's interference law, proposing  that the secondary wavelets mutually interfere. 
This combination, known as the Huygens-Fresnel principle,  
correctly describes the propagation of EM waves in free space under specific conditions \cite{Orfanidis}. 

The Rayleigh-Sommerfeld integral \cite{Orfanidis} mathematically describes the propagation of a wave
from one plane to another and is derived from the Huygens-Fresnel principle.
For simplicity, we focus on scalar fields, although this treatment can be extended to vector fields as well.
In a source-free region, where no charges or currents are present, the propagation of EM waves in a 
linear, isotropic, homogeneous, and nonconducting medium is governed by the {\em wave equation}
\be
\nabla^2 E(\bm r;t) = \frac{1}{c^2} \, \frac{\partial^2}{\partial t^2} E(\bm r;t)
\label{eq:wave-1}
\ee
where  
$E(\bm r;t)$ represents the electric field at an arbitrary point 
in space $\bm r \equiv (x,y,z)$ at  time $t$.

\begin{figure}[t]
\centering
\includegraphics[width=\linewidth]{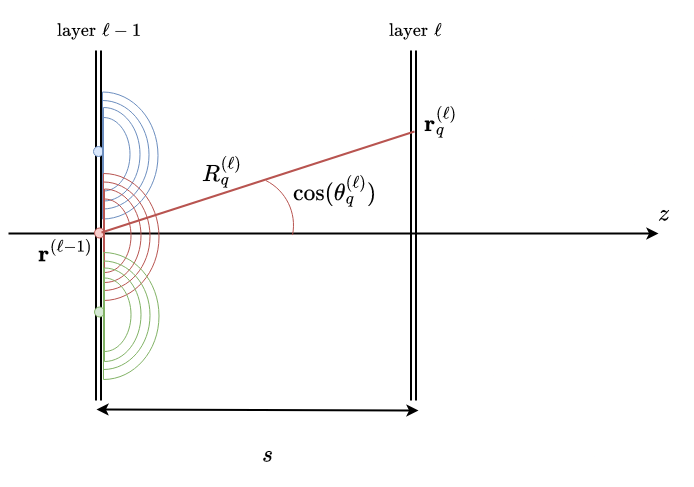} 
\caption{Field propagation from the $(\ell-1)$-th layer to the 
$\ell$-th one, $\ell \in \mathcal{L}-\{1\}$, based on the  Huygens-Fresnel principle.
When $\ell=1$, the $0$-th layer degenerates into the array of the BS (not reported here).
}
\label{fig:fig_13}
\end{figure}

\subsection{Monochromatic waves}

Let us consider a time-harmonic field with sinusoidal time-dependence given by 
\be
E(\bm r;t) = \Re \left\{u(\bm r) \, e^{j 2 \pi f_0 t}\right\}
\label{eq:wave-2}
\ee
where $f_0$ is the carrier frequency and $u(\bm r)$ is
a complex-valued field 
dependent on the spatial coordinates $\bm r$ but not on time $t$.
A wave of this form is known as a ``pure tone" or a ``monochromatic" signal.
Taking the Laplacian and the second time-derivative of \eqref{eq:wave-2}, we obtain from 
\eqref{eq:wave-1} that
the spatial part must satisfy the time-independent partial differential equation 
\be
\nabla^2 u(\bm r) + k_0^2 u(\bm r) = 0
\label{eq:wave-3}
\ee
where 
$k_0 = 2 \pi/\lambda_0$ is the wavenumber.
This equation is known as the {\em Helmholtz equation} \cite{Orfanidis}. 
Two possible solutions to the Helmholtz equation are plane waves and spherical waves.
Let us consider two adjacent layers $\ell-1$ and $\ell$ of the SIM,
as illustrated in Fig.~\ref{fig:fig_13}. Here, 
for $\ell \in \mathcal{L}-\{1\}$, 
$\br^{(\ell-1)} \equiv (x^{(\ell-1)}, y^{(\ell-1)},z^{(\ell-1)})$ identifies 
an arbitrary 
spatial point on the $(\ell-1)$-th layer, whereas the point
$\br_{q}^{(\ell)} \equiv (x_q^{(\ell)}, y_q^{(\ell)},z^{(\ell-1)}+s)$ specifies the location of the center of the 
$q$-th meta-atom in the $\ell$-th layer, for $q \in \mathcal{Q}$, and $s \in \Rset$ represents the distance 
between the two layers along the $z$-axis.
When $\ell=1$, the $0$-th layer corresponds to the antenna array of the BS.
In this case, $\br^{(0)} \equiv (x^{(0)}, y^{(0)},z^{(0)})$ represents  
an arbitrary  spatial point on the array, the vector 
$\br_{q}^{(1)} \equiv (x_q^{(1)}, y_q^{(1)},z^{(0)}+\sigma)$ denotes the location of the center of the 
$q$-th meta-atom in the first layer, for $q \in \mathcal{Q}$, and $\sigma$ is the distance 
between the antenna array of the BS and the first layer of the SIM.
In the following discussion, we focus on the {\em forward propagation} of the electric field.

For $\ell \in \mathcal{L}-\{1\}$, 
each point in the $(\ell-1)$-th layer 
emits a spherical wave ${e^{j k_0 R^{(\ell)}_q}}/{R^{(\ell)}_q}$. The amplitude of this wave
is multiplied by a directional factor
$s/R^{(\ell)}_q=\cos(\theta^{(\ell)}_q)$ (see Fig.~\ref{fig:fig_13}),
where 
\be
R_q^{(\ell)} = \|\br_{q}^{(\ell)} - \br^{(\ell-1)}\|
\ee
represents the distance between the source point on the 
$(\ell-1)$-th layer 
and the center of the $q$-th meta-atom of the adjacent layer.
The total field impinging on the $q$-th meta-atom of the $\ell$-th layer is given by 
the {\em Rayleigh-Sommerfeld} diffraction integral \cite{Orfanidis}:
\begin{multline}
u_{\text{in}, q}^{(\ell)} \eqdef u(\br_{q}^{(\ell)}) =  \iint \frac{s}{R^{(\ell)}_q} \left(\frac{1}{R^{(\ell)}_q} - j \, k_0\right) \,
\frac{e^{j k_0 R^{(\ell)}_q}}{2 \pi R^{(\ell)}_q} \\ \cdot  
u(\br^{(\ell-1)}) \, {\rm d}x^{(\ell-1)} \, {\rm d}y^{(\ell-1)}
\label{eq:diffr}
\end{multline}
for $\ell \in \mathcal{L}-\{1\}$.
When $\ell=1$, the integral \eqref{eq:diffr} still holds by
replacing $s$ with $\sigma$.
It should be noted that the propagator in \eqref{eq:diffr}
does not rely on any far-field assumption.
For the 
$q$-th meta-atom in the $\ell$-th layer,  the transmitted signal is obtained by multiplying
the incident signal by 
the transmission coefficient $\gamma_{\ell,q}(f_0)$.\footnote{In general, the transmission
coefficients are direction-dependent. However, we assume that 
the meta-atoms are judiciously engineered  to exhibit angular stability, meaning they have a weak dependence on the incidence direction. } 
This relationship is expressed as
\be
u_{\text{out}, q}^{(\ell)} = \gamma_{\ell,q}(f_0) \, u_{\text{in}, q}^{(\ell)}
\label{eq:trasmis}
\ee
for $\ell \in \mathcal{L}$ and $q \in \mathcal{Q}$. 
For a given metasurface, the transmission coefficient is typically calculated using unit-cell full-wave electromagnetic simulations, which assume local periodicity. This assumption is valid only if all higher-order grating modes are evanescent and sufficiently attenuated before reaching the adjacent metasurface, so that the near-field interactions can be neglected.
Although we did not focus on the specific implementation of the unit cells, we have estimated that, given the inter-element spacing and inter-layer separations used in our examples, these conditions are reasonably satisfied.

As previously mentioned, we assume that all layers are perfectly impedance matched, ensuring that multiple reflections are negligible. Consequently, the sole contribution to the forward propagation is given by the transmission.

To calculate the Rayleigh-Sommerfeld integral in a computationally effective fashion, we assume that,
the field on the $(\ell-1)$-th layer has a $2$-D discrete distribution given by
\begin{multline}
u(\br^{(\ell-1)}) = A_{\text{meta}} \sum_{\tilde{q}=0}^{Q-1} \, u(\br_{\tilde{q}}^{(\ell-1)}) \, \delta(x^{(\ell-1)}-x_{\tilde{q}}^{(\ell-1)}) 
\\ \cdot 
\delta(y^{(\ell-1)}-y_{\tilde{q}}^{(\ell-1)})
\label{eq:imp-field}
\end{multline}
for $\ell \in \mathcal{L}-\{1\}$, where the field is ideally concentrated at  
the centers of the meta-atoms in the $(\ell-1)$-th layer, each 
having a physical area of $A_{\text{meta}}$.
Similarly, in the case of $\ell=1$, we assume that
\begin{multline}
u(\br^{(0)}) = A_{\text{bs}}(f_0) \sum_{n=0}^{N-1} \, u(\br_{n}^{(0)}) \, \delta(x^{(0)}-x_{n}^{(0)}) 
\\ \cdot 
\delta(y^{(0)}-y_{n}^{(0)})
\label{eq:imp-field-0}
\end{multline}
where $\br_{n}^{(0)} \equiv (x_n^{(0)}, y_n^{(0)},z^{(0)})$,  
$A_{\text{bs}}(f_0)$, and $u(\br_{n}^{(0)})$ are 
the position, the effective area, and the baseband excitation signal emitted by 
the $n$-th antenna of the BS, for $n \in \mathcal{N}$.

By substituting \eqref{eq:imp-field} into \eqref{eq:diffr}
and invoking the sampling property of the Dirac delta, 
we obtain from \eqref{eq:trasmis} that 
\begin{multline}
u_{\text{in}, q}^{(\ell)} = \gamma_{\ell,q}(f_0) \, A_{\text{meta}} \sum_{\tilde{q}=0}^{Q-1} 
u(\br_{\tilde{q}}^{(\ell-1)}) \, \frac{s}{d^{(\ell)}_{q,\tilde{q}}} 
\\ \cdot \left(\frac{1}{d^{(\ell)}_{q,\tilde{q}}} - j \, k_0\right) \,
\frac{e^{j k_0 d^{(\ell)}_{q,\tilde{q}}}}{2 \pi d^{(\ell)}_{q,\tilde{q}}}
\label{eq:uq}
\end{multline}
for $\ell \in \mathcal{L}-\{1\}$, where $d^{(\ell)}_{q,\tilde{q}}$ is defined in \eqref{eq:dqq}.
Similarly, when $\ell=1$, by replacing $s$ with $\sigma$ in \eqref{eq:diffr}, 
we obtain from \eqref{eq:trasmis} and \eqref{eq:imp-field-0} that  
\begin{multline}
u_{\text{in}, q}^{(1)} = \gamma_{1,q}(f_0) \, A_{\text{bs}}(f_0) \sum_{n=0}^{N-1} \, u(\br_{n}^{(0)})
\, \frac{\sigma}{d^{(1)}_{q,n}} 
\\ \cdot \left(\frac{1}{d^{(1)}_{q,n}} - j \, k_0\right) \,
\frac{e^{j k_0 d^{(1)}_{q,n}}}{2 \pi d^{(1)}_{q,n}}
\label{eq:uq-1}
\end{multline}
where $d^{(1)}_{q,n}$ is defined in \eqref{eq:dqq-1}.
The model in \eqref{eq:forward} is directly derived from \eqref{eq:uq} and 
\eqref{eq:uq-1} after applying appropriate scaling to 
ensure the correct dimensions of the involved 
physical quantities.

\subsection{Multichromatic waves}

So far, we have considered strictly monochromatic waves. 
Now, let us examine the case where $E(\bm r;t)$ is a ``multichromatic" signal, defined as 
\be
E(\bm r;t) = \Re \left\{u(\bm r;t) \, e^{j 2 \pi f_0 t}\right\}
\label{eq:wave-multi}
\ee
In this context, the complex envelope $u(\bm r;t)$ has a Fourier transform (with respect to $t$), denoted by
\be
\EuScript{U}(\bm r;f) = \int_{-\infty}^{+\infty} u(\bm r;t) \, e^{-j 2 \pi f t} \, {\rm d}t
\ee
This transform assumes nonnegligible values over a spectral set $\mathcal{W}_u$ of nonzero 
measure $B_u$ centered around $f=0$.
For our purposes, we assume that $B_u \ll f_0$ ({\em narrowband assumption}).

The time-dependent field $u(\bm r;t)$ can be expressed as a continuous linear
combination of time-harmonic fields $\EuScript{U}(\bm r;f) \, e^{j 2 \pi f t}$ with varying 
frequencies $f \in \mathbb{R}$, given  by: 
\be
u(\bm r;t)  = \int_{\mathcal{W}_u} \EuScript{U}(\bm r;f) \, e^{j 2 \pi f t} \, {\rm d}f \: .
\label{eq:multi-1}
\ee
It can be shown that if $u(\bm r;t)$ is a solution of the wave equation \eqref{eq:wave-1}, 
then its Fourier transform $\EuScript{U}(\bm r;f)$ at a specific frequency $f$
is 
a solution to the Helmholtz equation \eqref{eq:wave-3}. This is done by replacing
$u(\bm r)$ with $\EuScript{U}(\bm r;f)$ and the wavenumber $k_0$
with $k(f) = 2 \pi f/c$.
Under appropriate regularity conditions,
the Rayleigh-Sommerfeld diffraction integral \eqref{eq:diffr} can be applied to each Fourier component $\EuScript{U}(\bm r;f)$ separately. This generalizes  
the transmitted signal by the $q$-th meta-atom of the $\ell$-th layer
given in \eqref{eq:trasmis} for the monochromatic case to the multichromatic case as follows 
\begin{multline}
u_{\text{out}, q}^{(\ell)}(t) = 
\int_{\mathcal{W}_u} \gamma_{\ell,q}(f)
\left\{ \iint \frac{s}{R^{(\ell)}_q} \, \left[\frac{1}{R^{(\ell)}_q} - j \, k(f) \right] 
\right. \\ \left. \cdot
\frac{e^{j k(f) R^{(\ell)}_q}}{2 \pi R^{(\ell)}_q} \,    
\EuScript{U}(\br^{(\ell-1)};f)  \, {\rm d}x^{(\ell-1)} \, {\rm d}y^{(\ell-1)}
\right\} e^{j 2 \pi f t} {\rm d}f
\label{eq:trasmis-multi}
\end{multline}
for $\ell \in \mathcal{L}$ and $q \in \mathcal{Q}$.
In general, the transmission coefficients of the SIM are 
frequency-dependent. However, for a narrowband signal, this
frequency dependence can be neglected by assuming that, within the spectral 
interval $\mathcal{W}_u$, the meta-atoms' transmission 
coefficients remain approximately consistent with their values at the carrier frequency $f_0$,
i.e., $\gamma_{\ell,q}(f) \approx \gamma_{\ell,q}(f_0)$.
Moreover, if $B_u \ll f_0$, we can further approximate $k(f) \approx k_0$.
Under these approximations, eq.~\eqref{eq:trasmis-multi} simplifies to 
\begin{multline}
u_{\text{out}, q}^{(\ell)}(t) \approx
\gamma_{\ell,q}(f_0)
\iint \frac{s}{R^{(\ell)}_q} \, \left[\frac{1}{R^{(\ell)}_q} - j \, k_0 \right] 
\\ \cdot
\frac{e^{j k_0 R^{(\ell)}_q}}{2 \pi R^{(\ell)}_q} \,    
u(\br^{(\ell-1)};t) \, {\rm d}x^{(\ell-1)} \, {\rm d}y^{(\ell-1)}
\label{eq:trasmis-multi-2}
\end{multline}
for $\ell \in \mathcal{L}$ and $q \in \mathcal{Q}$.
Starting from \eqref{eq:trasmis-multi-2}, we can derive the
model in \eqref{eq:forward} by reasoning similarly to
the previous subsection. This involves replacing 
$u(\br_{\tilde{q}}^{(\ell-1)})$ in \eqref{eq:imp-field}
and $u(\br_{n}^{(0)})$ in \eqref{eq:imp-field-0}
with $u(\br_{\tilde{q}}^{(\ell-1)}; t)$
and $u(\br_{n}^{(0)}; t)$, respectively.
Additionally, we assume that the transmit antennas of the BS
obey $A_{\text{bs}}(f) \approx A_{\text{bs}}(f_0)$.


\begin{IEEEbiography}[
{\includegraphics[width=1in,height=1.25in,clip,keepaspectratio]{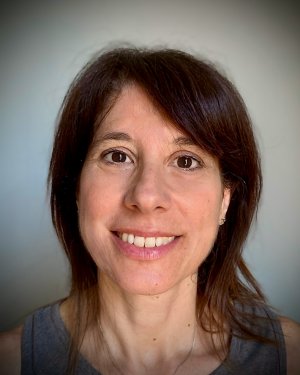}}]
{Donatella Darsena\,} (Senior Member, IEEE) received the Dr. Eng. degree summa cum laude in telecommunications engineering in 2001, and the Ph.D. degree in electronic and telecommunications engineering in 2005, both from the University of Napoli Federico II, Italy. From 2001 to 2002, she worked as embedded system designer in the Telecommunications, Peripherals and Automotive Group, STMicroelectronics, Milano, Italy. 
In 2005 she joined the Department of Engineering at Parthenope University of Napoli, Italy and worked first as an Assistant Professor and then as an Associate Professor from 2005 to 2022.
She is currently an Associate Professor in the Department of Electrical Engineering and Information Technology of the University of Napoli Federico II, Italy.
Her research interests are in the broad area of signal processing for communications, with current emphasis on reflected-power communications, orthogonal and nonorthogonal multiple access techniques, wireless system optimization, and physical-layer security.
Dr. Darsena has served as a Senior Editor for IEEE ACCESS since 2024, Executive Editor for IEEE COMMUNICATIONS LETTERS since 2023, and Associate Editor for IEEE SIGNAL PROCESSING LETTERS since 2020. She was an Associate Editor of IEEE ACCESS (from 2018 to 2023), of IEEE COMMUNICATIONS LETTERS (from 2016 to 2019), and Senior Area Editor of IEEE COMMUNICATIONS LETTERS (from 2020 to 2023). 
\end{IEEEbiography}

\begin{IEEEbiography}[
{\includegraphics[width=1in,height=1.25in,clip,keepaspectratio]{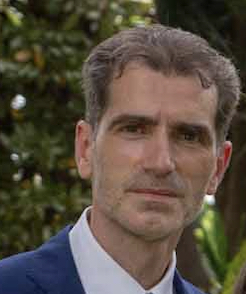}}]
{Francesco Verde\,} (Senior Member, IEEE)  received the Dr. Eng. degree
\textit{summa cum laude} in electronic engineering
from the Second University of Napoli, Italy, in 1998, and the Ph.D.
degree in information engineering
from the University of Napoli Federico II, in 2002.
Since December 2002, he has been with the University of Napoli Federico II, Italy. He first served as an Assistant Professor of signal theory and mobile communications
and, since December 2011, he has served as an Associate Professor of telecommunications with the Department of Electrical Engineering and Information Technology.
His research activities include reflected-power communications, 
orthogonal/non-orthogonal multiple-access techniques, wireless systems optimization, and 
physical-layer security.

Prof. Verde has been involved in several technical program committees of major IEEE conferences in signal processing and wireless communications.
He has served as Associate Editor for IEEE TRAN\-SACTIONS ON VEHICULAR TECHNOLOGY since 2022.
He was an Associate Editor of the IEEE TRANSACTIONS ON SIGNAL PROCESSING (from 2010 to 2014), IEEE SIGNAL PROCESSING LETTERS (from 2014 to 2018),
IEEE TRANSACTIONS ON COMMUNICATIONS (from 2017 to 2022), and 
Senior Area Editor of the IEEE SIGNAL PROCESSING LETTERS (from 2018 to 2023), 
as well as Guest Editor of the EURASIP Journal on Advances in Signal Processing in 2010 and SENSORS MDPI in 2018-2022.
\end{IEEEbiography}

\begin{IEEEbiography}
[{\includegraphics[width=1in,height=1.25in,clip,keepaspectratio]{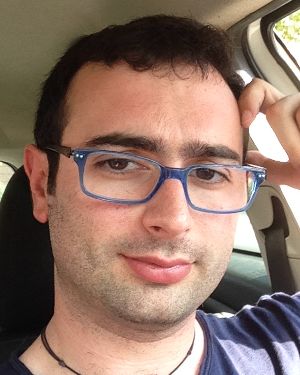}}]
{Ivan Iudice\,} received the B.S. and M.S. degrees
in telecommunications engineering in 2008 and 2010,
respectively, and the Ph.D. degree
in information technology and electrical engineering in 2017,
all from University of Napoli Federico II, Italy.

Since November 2011,
he has been with the Italian Aerospace Research Centre (CIRA), Capua, Italy.
He first served as part of the Electronics and Communications Laboratory
and he is currently part of the Security Unit.
He is involved in several international projects.
He serves as reviewer for several international journals
and as TPC member for several international conferences.
He is author of several papers on refereed journals and international conferences.
His research activities mainly lie in the area of
signal and array processing for communications,
with current interests focused on physical-layer security,
space-time techniques for cooperative communications systems
and reconfigurable metasurfaces.
\end{IEEEbiography}

\begin{IEEEbiography}[{\includegraphics[width=1in,height=1.25in,clip,keepaspectratio]{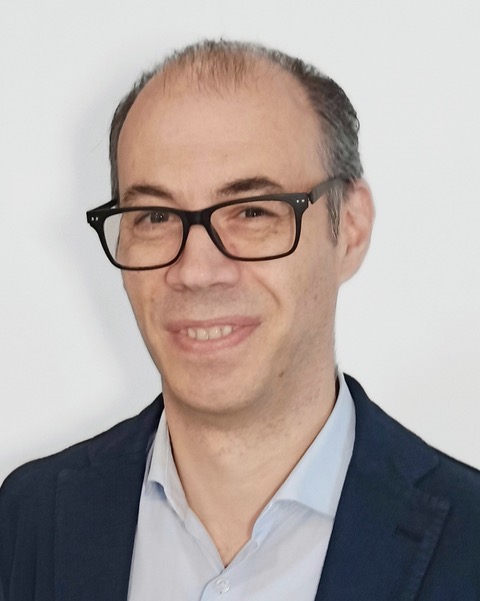}}]
	{Vincenzo Galdi\,} (Fellow, IEEE) 
	received the Laurea degree ({\em summa cum laude}) in electrical engineering and the Ph.D. degree in applied electromagnetics from the University of Salerno, Italy, in 1995 and 1999, respectively. 
	
	He has held several research-associate and visiting positions at abroad research institutions, including the European Space Research and Technology Centre, Noordwijk, The Netherlands; Boston University, Boston, MA, USA; the Massachusetts Institute of Technology, Cambridge, MA, USA; the California Institute of Technology, Pasadena, CA, USA; and The University of Texas at Austin, Austin, TX, USA. He is currently a Professor of electromagnetics with the Department of Engineering, University of Sannio, Benevento, Italy, where he leads the Fields \& Waves Laboratory. He is the Co-Founder of the spinoff company MANTID srl, Benevento, and the startup company BioTag srl, Naples. He has co-edited two books and coauthored about 180 articles in peer-reviewed international journals, and is the co-inventor of thirteen patents. His research interests encompass wave interactions with complex structures and media, multiphysics metamaterials, smart propagation environments, optical sensing, and gravitational interferometry. 
	
	Dr. Galdi is a Fellow of Optica (formerly OSA), a Senior Member of the LIGO Scientific Collaboration, and a member of the American Physical Society. He was a recipient of the Outstanding Associate Editor Award of \textsc{IEEE Transactions on Antennas and Propagation} in 2014 and the URSI Young Scientist Award in 2001. He has served as the Chair for the Technical Program Committee of the International Congress on Engineered Material Platforms for Novel Wave Phenomena in 2018, a Topical/Track Chair for the Technical Program Committee of the IEEE International Symposium on Antennas and Propagation and USNC-URSI Radio Science Meeting from 2016 to 2017 and from 2020 to 2023, and an organizer/chair for several topical workshops and special sessions. He has also served as a Track Editor from 2016 to 2020, a Senior Associate Editor from 2015 to 2016, and an Associate Editor from 2013 to 2014 of the \textsc{IEEE Transactions on Antennas and Propagation}. He is serving as an Associate Editor of {\em Optics Express} and a regular reviewer for many journals, conferences, and funding agencies.	
\end{IEEEbiography}

\end{document}